\documentclass[twocolumn]{aastex61}

\let\svthefootnote\thefootnote

\usepackage{multirow}
\usepackage{amsmath}

\def\iso#1{$^{#1}$}

\submitjournal{ApJ}

\shorttitle{Stochastic Evolution of Radioactive Isotopes}
\shortauthors{C\^ot\'e et al.}

\begin{document}

\title{Stochastic Chemical Evolution of Radioactive Isotopes \\ with a Monte Carlo Approach}

\correspondingauthor{Benoit C\^ot\'e}
\email{benoit.cote@csfk.mta.hu}

\author[0000-0002-9986-8816]{Benoit C\^ot\'e$^*$}
\affiliation{Konkoly Observatory, Research Centre for Astronomy and Earth Sciences, MTA Centre for Excellence, Konkoly Thege Mikl\'os \'ut 15-17, H-1121 Budapest, Hungary}
\affiliation{National Superconducting Cyclotron Laboratory, Michigan State University, East Lansing, MI 48824, USA}
\affiliation{Joint Institute for Nuclear Astrophysics - Center for the Evolution of the Elements, USA}
\affiliation{NuGrid Collaboration, \url{http://nugrid.github.io/}}
 
\author[0000-0002-7294-9288]{Andr\'es Yag\"ue$^*$}
\affiliation{Konkoly Observatory, Research Centre for Astronomy and Earth Sciences, MTA Centre for Excellence, Konkoly Thege Mikl\'os \'ut 15-17, H-1121 Budapest, Hungary}

\author{Blanka Vil\'agos}
\affiliation{Konkoly Observatory, Research Centre for Astronomy and Earth Sciences, MTA Centre for Excellence, Konkoly Thege Mikl\'os \'ut 15-17, H-1121 Budapest, Hungary}

\author{Maria Lugaro}
\affiliation{Konkoly Observatory, Research Centre for Astronomy and Earth Sciences, MTA Centre for Excellence, Konkoly Thege Mikl\'os \'ut 15-17, H-1121 Budapest, Hungary}
\affiliation{Monash Centre for Astrophysics, School of Physics and Astronomy, Monash University, VIC 3800, Australia}\let\thefootnote\relax\footnotetext{$^*$ These authors put equal amount of effort on the numerical development and on the writing of the manuscript.}\addtocounter{footnote}{-1}\let\thefootnote\svthefootnote

%\let\thefootnote\relax\footnotetext{$^*$ The authors put equal amount of effort on the numerical development and on the writting of the manuscript.}

%% Mark off the abstract in the ``abstract'' environment. 
\begin{abstract}

Short-lived radionuclides (SLRs) with mean-lives $\tau$ of a few to hundreds Myr provide unique opportunities to probe recent nucleosynthesis events in the interstellar medium, and the physical conditions in which the Sun formed. Here we quantify the uncertainty in the predicted evolution of SLRs within a parcel of interstellar gas given the stochastic nature of stellar enrichment events. We assume that an enrichment progenitor is formed at every time interval $\gamma$. For each progenitor, we randomly sample the delay time between its formation and its enrichment event, based on several delay-time distribution (DTD) functions that cover a wide range of astrophysical sites. For each set of $\tau$, $\gamma$, and DTD function, we follow the abundances of SLRs for 15\,Gyr, and repeat this process thousands of times to derive their probability distributions. For $\tau/\gamma\gtrsim2$, the distributions depend on the DTD function and we provide tabulated values and analytical expressions to quantify the spread. The relative abundance uncertainty reaches a maximum of $\sim$\,60\,\% for $\tau/\gamma=1$. For $\tau/\gamma\lesssim1$, we provide the probability for the SLR abundance to carry the signature of only one enrichment event, which is greater than 50\,\% when $\tau/\gamma\lesssim0.3$. For $0.3\lesssim\tau/\gamma\lesssim 2$, a small number of events contributed to the SLR abundance. This case needs to be investigated with a separate statistical method. We find that an isolation time for the birth of the Sun of roughly $9-13$\,Myr is consistent with the observed abundances of \iso{60}Fe, \iso{107}Pd, and \iso{182}Hf in the early Solar System, when assuming $\tau/\gamma\sim3$ for these isotopes.

\end{abstract}

%% Keywords should appear after the \end{abstract} command. 
%% See the online documentation for the full list of available subject
%% keywords and the rules for their use.
\keywords{ISM: abundances -- meteorites, meteors, meteoroids -- planets and satellites: formation}

%%%%%%%%%%%%%%%%%%%%%
%%%%%%%%%%%%%%%%%%%%%
%%%%%%%%%%%%%%%%%%%%%
\section{Introduction}
\label{sec:intro}
%%%%%%%%%%%%%%%%%%%%%
%%%%%%%%%%%%%%%%%%%%%
%%%%%%%%%%%%%%%%%%%%%

Radioactive nuclei with half-lives of the order of a few tens to a few hundreds Myr (short-lived radionuclides, SLRs hereafter) are one of the most promising tools to investigate a variety of current astrophysical topics including ongoing stellar nucleosynthesis \citep{radiobook2}, the properties of different  rapid neutron-capture ($r$) process sites \citep{hotokezaka15}, the physics of the interstellar medium (ISM) \citep{krause18} and of the local bubble \citep{feige18}, the formation of the Solar System \citep{Lugaro2018}, and the thermo-mechanical evolution of protoplanets \citep{lichtenberg19}. The peculiarity of radioactive nuclei is that they decay into a daughter nucleus with a characteristic timescale determined by their half-life. This means they can be used as fingerprints of current nucleosynthesis \citep{diehl13} and of the dynamics of massive star associations and star-forming regions \citep{diehl10}, as well as cosmochronometers \citep{lugaro14,lugaro16}. Their decay also need to be considered as a heating factor in protoplanetary disks \citep{lichtenberg16b}.

Modeling the evolution of SLR abundances in the ISM is however complicated by their decaying nature. First, these nuclei are affected differently than stable nuclei by the star formation history, the gas circulation processes (inflows and outflows) occurring within the Galaxy, and the delay time\footnote{Throughout the paper we will use the term $delay\,\,time$ to indicate the time that elapses between the formation of a progenitor stellar source and the associated enrichment event, for example, between the formation of a massive star and its final core-collapse supernova.} of different stellar sources \citep{clayton85,huss09,cote19}. For example, SLRs are more sensitive to the star formation rate at any given time, while stable nuclei are more sensitive to the total integrated star formation history. By considering these effects and their uncertainties, \citet[hereafter Paper~I]{cote19} proposed a multiplication factor of $2.3^{+3.4}_{-0.7}$ to the simple steady-state formula for the ratio between a radioactive and stable nucleus (see their Equation 1). They also analysed the limitations of using the steady-state formula and provided an open-source code to calculate the evolution of any radioactive nucleus in the Galaxy using mass- and metallicity-dependent stellar yields and proper delay-time distribution functions for different stellar sources. These results have allowed us to quantify the global uncertainties on the predicted radio-to-stable isotopic ratios in the ISM. However, in Paper~I, we did not consider the heterogeneities in the ISM due to the fact that stellar enrichment events are discrete both in time and space. These heterogeneities, coupled with the fact that radioactive nuclei decay,  can result in large SLR abundance fluctuations in different parcels of gas within the Galaxy \citep{meyer00,wasserburg06}.

\begin{deluxetable*}{ll}
\tablewidth{0pc}
\tablecaption{Definition of the key terms involved in our quantification of radioactive isotope abundances uncertainties.\label{tab:def}}
\tablehead{\colhead{Term} & \colhead{Definition}
 }
\startdata
$\tau$ & Mean-life of a radioactive isotope. \\
$\gamma$ & Constant time interval between the formation of two progenitors. \\
$t_\mathrm{delay}$ & Time interval between the formation of a progenitor and its associated enrichment event. \\
$\delta$ & Time interval between two consecutive enrichment events. \\
$M_\mathrm{radio}$ & Mass of radioactive isotopes present in a given parcel of gas. \\
$N_\mathrm{SLR}$ & Number of radioactive isotopes, related to the mass by $M_\mathrm{radio}$ = $m_\mathrm{iso}N_\mathrm{SLR}$, where $m_\mathrm{iso}$ is the mass of the isotope. \\
Spread & 68\,\% confidence level interval, centered around the median, of a symmetric or non-symmetric $M_\mathrm{radio}$ distribution. \\
$\sigma$ & Standard deviation ($\sqrt{\langle M_\mathrm{radio}^2\rangle-\langle M_\mathrm{radio}\rangle^2}$). Half of the 68\,\% confidence level interval when $M_\mathrm{radio}$ is a \\
 & normal distribution. \\
$T_\mathrm{iso}$ & Isolation time of the molecular cloud in which the Sun was born. \\
\enddata
\end{deluxetable*}

3D hydrodynamical simulations have been used to follow the heterogeneous evolution of one or two selected SLRs, \iso{26}Al (half-life = 0.72 Myr) and \iso{60}Fe (2.62 Myr), in giant molecular clouds \citep{vasileiadis13}, star forming regions \citep{krause18}, and the ISM \citep{fujimoto18}. While these studies have provided the first best attempts at the problem of heterogeneous evolution of SLRs, their results should be treated with caution because the stellar production of \iso{26}Al and \iso{60}Fe from massive stars winds and core-collapse supernovae, one of the main inputs in these simulations, is still extremely uncertain \citep[see, e.g.][]{limongi06,iliadis11,austin17,sukhbold16,jones19}. The problem of Galactic chemical evolution (GCE) heterogeneity was considered by \citet{hotokezaka15} and \cite{bartos19} using a Monte Carlo approach and assuming diffusive mixing in the ISM for $r$-process SLRs such as \iso{244}Pu (see also \citealt{tsujimoto17}). Many more SLRs and astrophysical sites are of interest and need to be investigated \citep[see Tables 1 and 2 of][]{Lugaro2018}. 

The aim of the present paper is to analyse heterogeneities of SLRs abundances in the ISM within a general statistical framework. We vary two crucial parameters that control the heterogeneous evolution of radioactive nuclei in a given parcel of gas within the Galaxy: (1) the mean-life $\tau$ of the considered SLR, which is linked to the half-life $T_{1/2}$ via $\tau=T_{1/2} / \ln 2$, and (2) the recurrent time $\gamma$ between the formation of progenitor stellar sources. The delay times $\delta$ between the different enrichment events that pollute the parcel of gas are defined by randomly sampling the delay-time distribution function of the stellar source, using a Monte Carlo approach (see Section~\ref{sec:mc}). Our simple approach has the advantage of being easily extendable to any SLR and astrophysical site of interest. 

Our results can be applied to any abundance ratio between a radioactive and a stable nucleus. In a forthcoming paper, we will apply the same analysis to ratios between two different radioactive nuclei (A. Yag\"ue, in preparation), which is particular important for the \iso{60}Fe/\iso{26}Al ratio and for ratios involving the actinides. Our work currently assumes that every enrichment event pollutes the same parcel of interstellar matter with the same amount of ejecta. Proper treatments of physical mechanisms that transport and mix SLR nuclei in the ISM in 3D will also be addressed in forthcoming papers. The present paper is the first step towards a better prediction of the complex physics of SLRs in the ISM.% against observational constraints.

The outline of this paper is as follows. In Section~\ref{sec:intro_sub} we review the basic equations describing the evolution of SLRs in a heterogeneous ISM. In Section~\ref{sec:mc} we describe our Monte Carlo setup designed to expand upon those equations and to quantify the uncertainties in the predicted abundances of SLRs. In Section~\ref{sec:delta_distribution} we present the distribution of time intervals elapsed between two consecutive enrichment events, assuming different delay-time distribution functions for the stellar source. In Section~\ref{sec:spread} we quantify the spread (or uncertainty) in the predicted abundances of SLRs in the ISM under different regimes, based on the mean-life of the considered SLR and the rarity of the stellar source. We discuss our results and apply them to the radioactive composition of the early Solar System in Section~\ref{sec:discusion}. We conclude in Section~\ref{sec:conclusion}.

%%%%%%%%%%%%%%%%%%%%%%%%%%%%%%%%%%%%%%
%%%%%%%%%%%%%%%%%%%%%%%%%%%%%%%%%%%%%%
\subsection{Analytical Basics of the Evolution of Radioactive Isotopes in a Heterogeneous Galaxy}
\label{sec:intro_sub}
%%%%%%%%%%%%%%%%%%%%%%%%%%%%%%%%%%%%%%
%%%%%%%%%%%%%%%%%%%%%%%%%%%%%%%%%%%%%%
The evolution of a SLR with mean-life $\tau$ in a parcel of interstellar matter can be mathematically expressed as the sum of the contribution of the $n$ enrichment events that contributed to such parcel of interstellar matter, each of the contributions decayed to the present time $t$ as
\begin{equation}
    N_\mathrm{SLR}(t) = e^{-(t - t_0)/\tau} + e^{-(t - t_1)/\tau} + ... + e^{- (t - t_n)/\tau},
    \label{eq:evRadioFirst}
\end{equation}
where $t_i$ represents the time at which the $i^{\mathrm{th}}$ enrichment event has taken place. By setting $t_0$ = 0 and defining $\delta_i = t_{i + 1} - t_i$ and $t_j = \sum_{i = 0}^{j - 1}\delta_i$, we can rewrite Equation~(\ref{eq:evRadioFirst}) as
\begin{equation}
    N_\mathrm{SLR}(t) = e^{-t/\tau} + \sum_{j = 0}^{n - 1} e^{-(t - \sum_{i = 0}^j\delta_i)/\tau}.
    \label{eq:evRadio}
\end{equation}
When taking $\delta_i$ as a constant, $\delta_c$, we recover the analysis presented in \cite{Lugaro2018}, which leads to
\begin{equation}
    N_\mathrm{SLR}(t) = \frac{1 - e^{-(n + 1)\delta_c/\tau}}{1 - e^{-\delta_c/\tau}}e^{-(t - n\delta_c)/\tau}.
    \label{eq:constDeltModel}
\end{equation}

\begin{figure*}
    %\center
\includegraphics[width=7.0in]{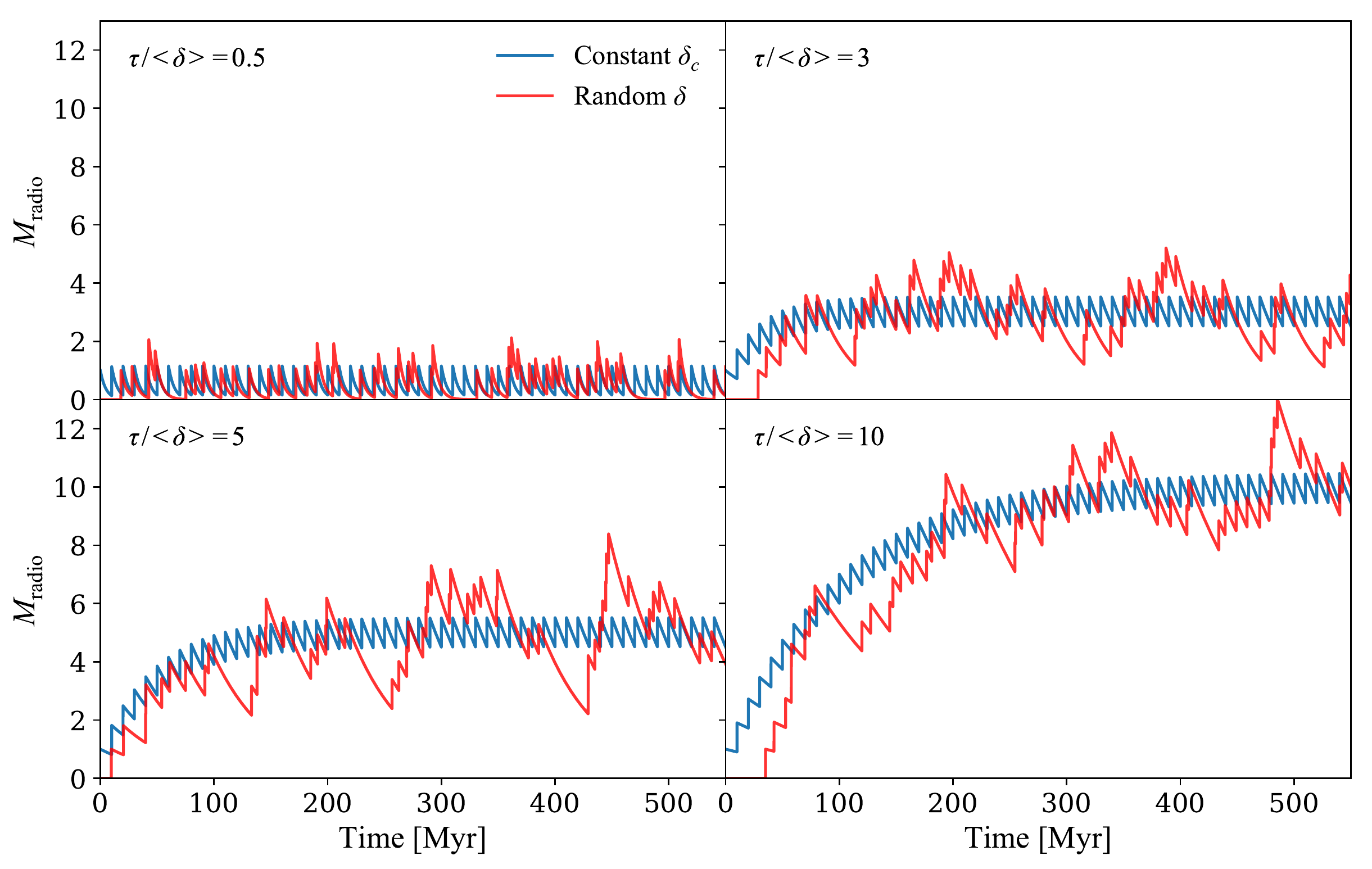}
    \caption{Evolution of the mass of radioactive isotopes found in a parcel of interstellar matter, assuming different ratios (different panels) between the mean-life of the isotopes $\tau$ and the average time interval $<\delta>$ between two successive enrichment events. The blue and red lines show the evolution when $\delta$ is assumed to be constant and random, respectively. In this figure, and throughout of this paper, each enrichment event ejects 1\,M$_\odot$ of radioactive material.
    }
    \label{fig:cte_vs_frm}
\end{figure*}

When all enrichment events are separated by a constant time interval $\delta_c$, as shown by the blue lines in Figure~\ref{fig:cte_vs_frm}, the evolution of the mass of radionuclei\footnote{The mass of radionuclei present in the ISM is simply $N_\mathrm{SLR}$ times the mass of a single isotope.} oscillates regularly and reaches a steady state after $n\sim5\,\tau/\delta_c$ events. The average value of this steady state is proportional to $\tau/\delta_c$ and the relative uncertainty around this value is $\delta_c/\tau$. The larger is $\tau$ compared to $\delta_c$, the less the SLR decays before the next enrichment event (see also Section~4.1 in \citealt{Lugaro2018}). On the other hand, if the time intervals between two consecutive events are defined randomly, the oscillations become stochastic and the abundances show stronger variations (red lines in Figure~\ref{fig:cte_vs_frm}). In that case, several events can pile up during a short period of time, or no events may occur for a relatively long period of time. This increases and reduces the maximum and minimum abundance values the system can reach, compared to the case with constant $\delta_c$.  The range of possible abundance values (i.e., the uncertainty) therefore grows substantially when considering $\delta$ as a random variable instead of a constant.  Because a random $\delta$ does not allow for a closed expression such as that of Equation~(\ref{eq:constDeltModel}), we study the properties of Equation~(\ref{eq:evRadio}) and use Monte Carlo calculations to generate the distribution of $\delta$.

%%%%%%%%%%%%%%%%%%%%%%%%%%%%%%%%%%
%%%%%%%%%%%%%%%%%%%%%%%%%%%%%%%%%%
%%%%%%%%%%%%%%%%%%%%%%%%%%%%%%%%%%
\section{Monte Carlo Calculations}
\label{sec:mc}
%%%%%%%%%%%%%%%%%%%%%%%%%%%%%%%%%%
%%%%%%%%%%%%%%%%%%%%%%%%%%%%%%%%%%
%%%%%%%%%%%%%%%%%%%%%%%%%%%%%%%%%%
Our Monte Carlo setup is designed to follow numerically the stochastic evolution of SLRs in the ISM. Our goal is to quantify the spread in the predicted SLR abundances in a simplified environment, where every enrichment event pollutes the same cloud of gas with the same amount of radionuclei. The purpose of this simplified framework is to isolate and focus on the role of the mean-life $\tau$ and the properties of enrichment events on the predicted spread. Throughout this paper, all events eject 1\,M$_\odot$ of a SLR isotope and 1\,M$_\odot$ of a stable reference isotope so that our results, including the spreads, can be normalized using actual yields and production ratios (as we will present as example in Section~\ref{sec:scenario}).

\begin{figure*}
%\center
\includegraphics[width=7.05in]{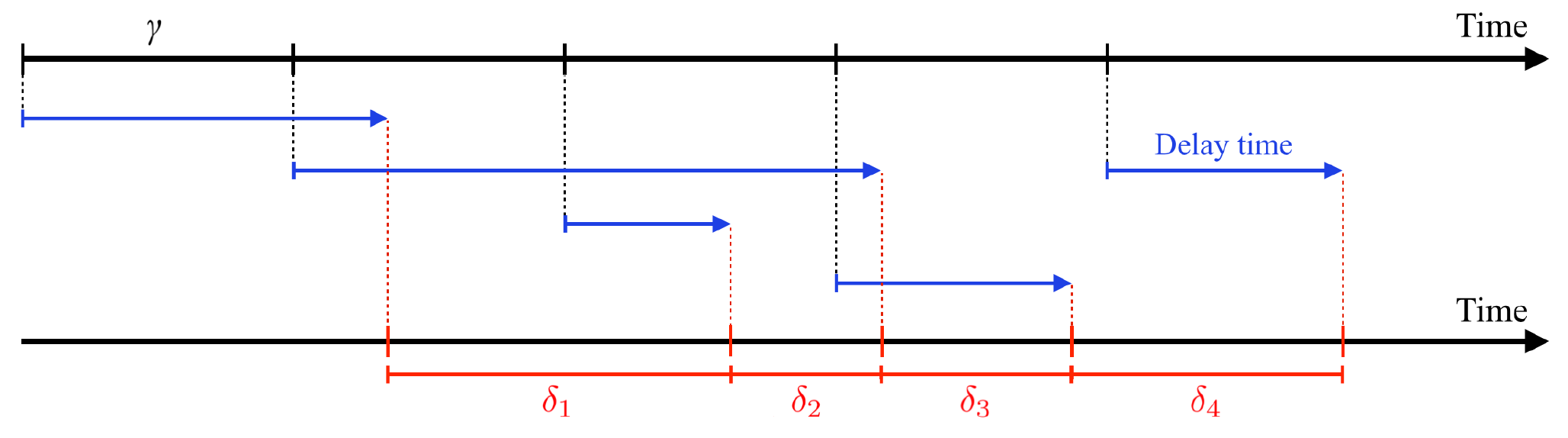}
\caption{Visualization of the key parameters involved in our Monte Carlo calculations. $\gamma$ is the constant time interval between the formation of two progenitors. The delay times (blue arrows) represent the time intervals between the formation of the progenitors and their associated enrichment event. Those delay times are randomly sampled from an input DTD function. The different $\delta$ values are the time intervals between two consecutive enrichment events, regardless of the formation time of the progenitors. This means that $\delta$ cannot have negative values.
}
\label{fig:param}
\end{figure*}

The stochastic study requires the introduction of the parameter $\gamma$ that represents the separation in time between the formation of the progenitor stars for the enrichment events. Throughout this work, we assume that this parameter is a constant, and that for each progenitor the enrichment event occurs after a certain delay time drawn randomly from a DTD function. The value of each $\delta$ is then calculated by taking two consecutive enrichment events (Figure~\ref{fig:param}). This means that even if the progenitors are assumed to form at a regular rate, the abundance of the SLR will be shed into the ISM stochastically. In the extreme case where all delay times are the same, then we have that $\delta = \gamma = \delta_c$ and the evolution of the SLR can be followed by using Equation~(\ref{eq:constDeltModel}). By assuming that $\gamma$ is a constant, we are implying a constant star formation history around a given parcel of gas. In future studies, we will explore the impact of adopting time-dependent star formation histories, which will make $\gamma$ a time-dependent variable. This could be important when the adopted DTD function covers a time window of several Gyr. Table~\ref{tab:def} lists the definition of all terms relevant for our study.
\\
\\
%%%%%%%%%%%%%%%%%%%%%%%%%%%%%%%%%%%%%%%%%%%%%%%%
%%%%%%%%%%%%%%%%%%%%%%%%%%%%%%%%%%%%%%%%%%%%%%%%
\subsection{Range of Mean-Lives and DTD Functions}
\label{sec:dtd_meanlives}
%%%%%%%%%%%%%%%%%%%%%%%%%%%%%%%%%%%%%%%%%%%%%%%%
%%%%%%%%%%%%%%%%%%%%%%%%%%%%%%%%%%%%%%%%%%%%%%%%
Our main goal is to characterize the spread of SLR abundances in a general way, so that our results can thereafter be applied to specific isotopes and known enrichment sources. For this reason, we consider a wide range of mean-lives that encompasses most the  radioactive isotopes that can be measured from meteorite data for the early Solar System (see Table~2 in \citealt{Lugaro2018}). However, to use our statistical framework (see Section~\ref{sec:convergence}), the radioactive abundances need to reach a steady state, or be near to a steady state, by the time the Solar System forms. Therefore, we exclude mean-lives of several Gyr, and rather focus on mean-lives between 1\,Myr and 1\,Gyr. This range excludes the longer-lived U and Th isotopes, as well as the shorter-lived, \iso{36}Cl and \iso{41}Ca. We will briefly discuss at the end of Section~\ref{sec:scenario} where these isotopes fit within our results. For the time interval between the formation of the progenitors, $\gamma$, we explore values between 1 and 316\,Myr, so that the ratio $\tau/\gamma$ mostly ranges from 0.01 to 316. Although $\gamma$ is not easy to determine from first principles, the range explored in this work encompasses the values estimated in \cite{Lugaro2018} for various enrichment sources (see also \citealt{meyer00} and \citealt{hotokezaka15}). 

We explore six DTD functions that are split into three categories: short-duration, medium-duration, and long-duration functions, spanning from 3\,Myr to 50\,Myr, from 50\,Myr to 1\,Gyr, and from 50\,Myr to 10\,Gyr, respectively. For each category, we explore a power-law DTD function in the form of $t^{-1}$ and a uniform ``box'' distribution where all delay times within the considered time frame have equal probability of being randomly picked. For an application, short-duration functions could be associated with core-collapse supernovae, while long-duration functions in the form of $t^{-1}$ could be associated with Type~Ia supernovae (e.g., \citealt{2009ApJ...699.2026R,2014ARA&A..52..107M}) and neutron star mergers (e.g., \citealt{2012ApJ...759...52D,2017ApJ...848L..23F,2018MNRAS.474.2937C}). The DTD of asymptotic giant branch (AGB) stars producing slow neutron-capture ($s$) process isotopes in the mass range roughly between 2 and 4\,M$_\odot$ (e.g., \citealt{lugaro14}) spans approximately 3\,Gyr and would therefore fall between a medium- and long-duration DTD function.

To quantify the spread in the mass $M_\mathrm{radio}$ of a SLR in the ISM as a function of time, we run multiple simulations for any given DTD function, $\gamma$, and $\tau$, where we randomly sample the delay time of each enrichment event. Each individual run (e.g., the red lines in Figure~\ref{fig:cte_vs_frm}) has its own unique temporal profile. To quantify the distribution of $M_\mathrm{radio}$, we stack all the runs together and calculate the median value of all predictions and the 68\,\% and 95\,\% confidence levels centered around that median, using time bins of 1\,Myr. Every simulation lasts for 15\,Gyr in total. Although delay times are randomly sampled, the evolution of $M_\mathrm{radio}$ is calculated analytically using Equation~(\ref{eq:evRadio}) from an input list of times at which events are occurring. This makes the evolution of $M_\mathrm{radio}$ independent of the adopted time resolution.

\begin{figure*}
%\center
\includegraphics[width=7.05in]{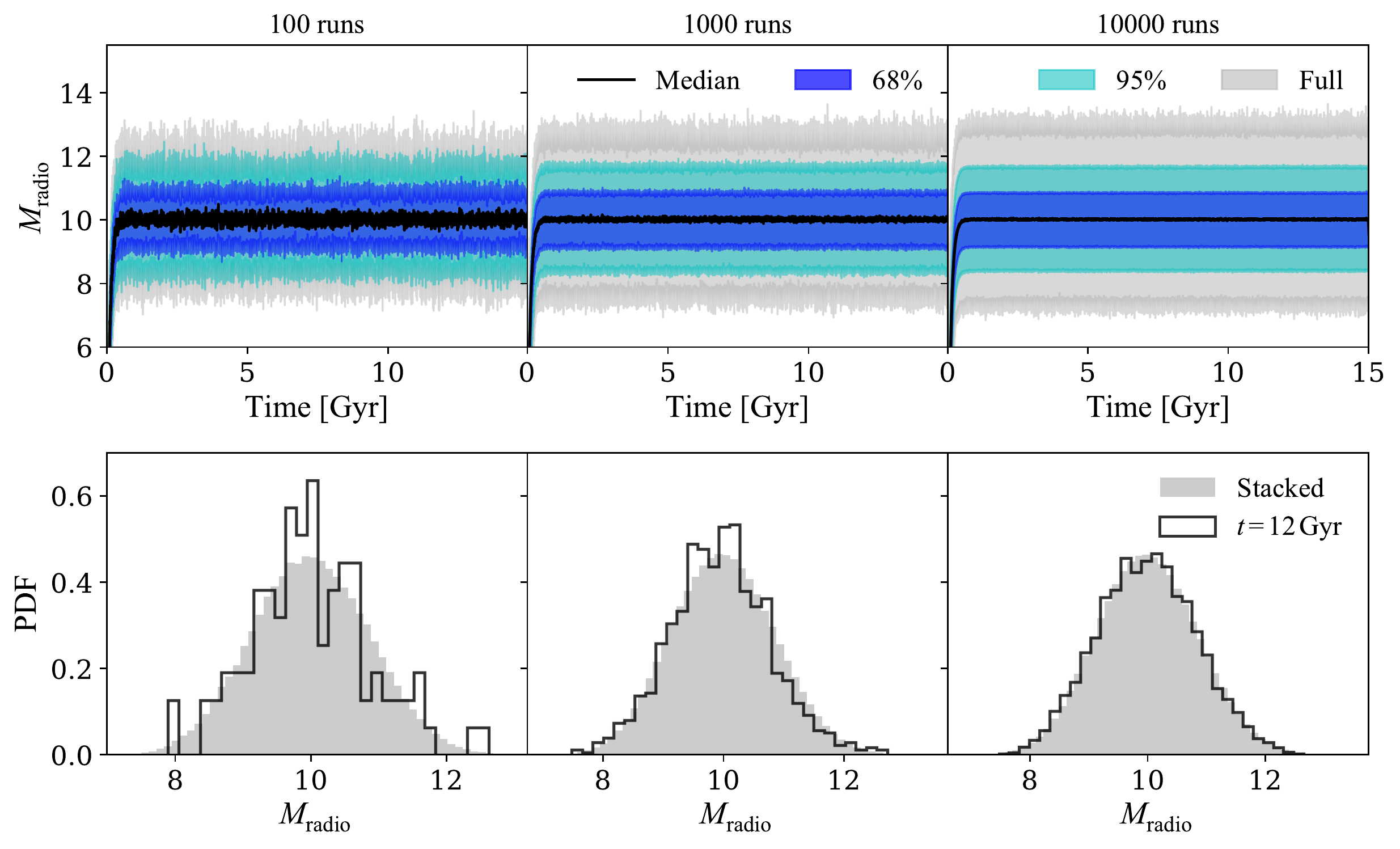}
\caption{Top panels: Evolution of the median and the 68\,\% and 95\,\% confidence levels of the mass of radionuclei ($M_\mathrm{radio}$) as a function of time, using Monte Carlo calculations with 100, 1000, and 10000 runs (from left to right), with $\gamma=10$\,Myr and the $3-50$\,Myr box DTD function. The grey shaded area represents the maximum and minimum values reached during the calculations. Bottom panels: Distribution of predicted $M_\mathrm{radio}$ at 12\,Gyr (empty histograms) for the calculations with 100, 1000, and 10000 runs (from left to right). The grey filled histograms show the distribution of $M_\mathrm{radio}$ when all timesteps between 12 and 14\,Gyr are stacked together to improve the statistics.}
\label{fig:convergence}
\end{figure*}

\begin{figure*}
%\center
\includegraphics[width=7.1in]{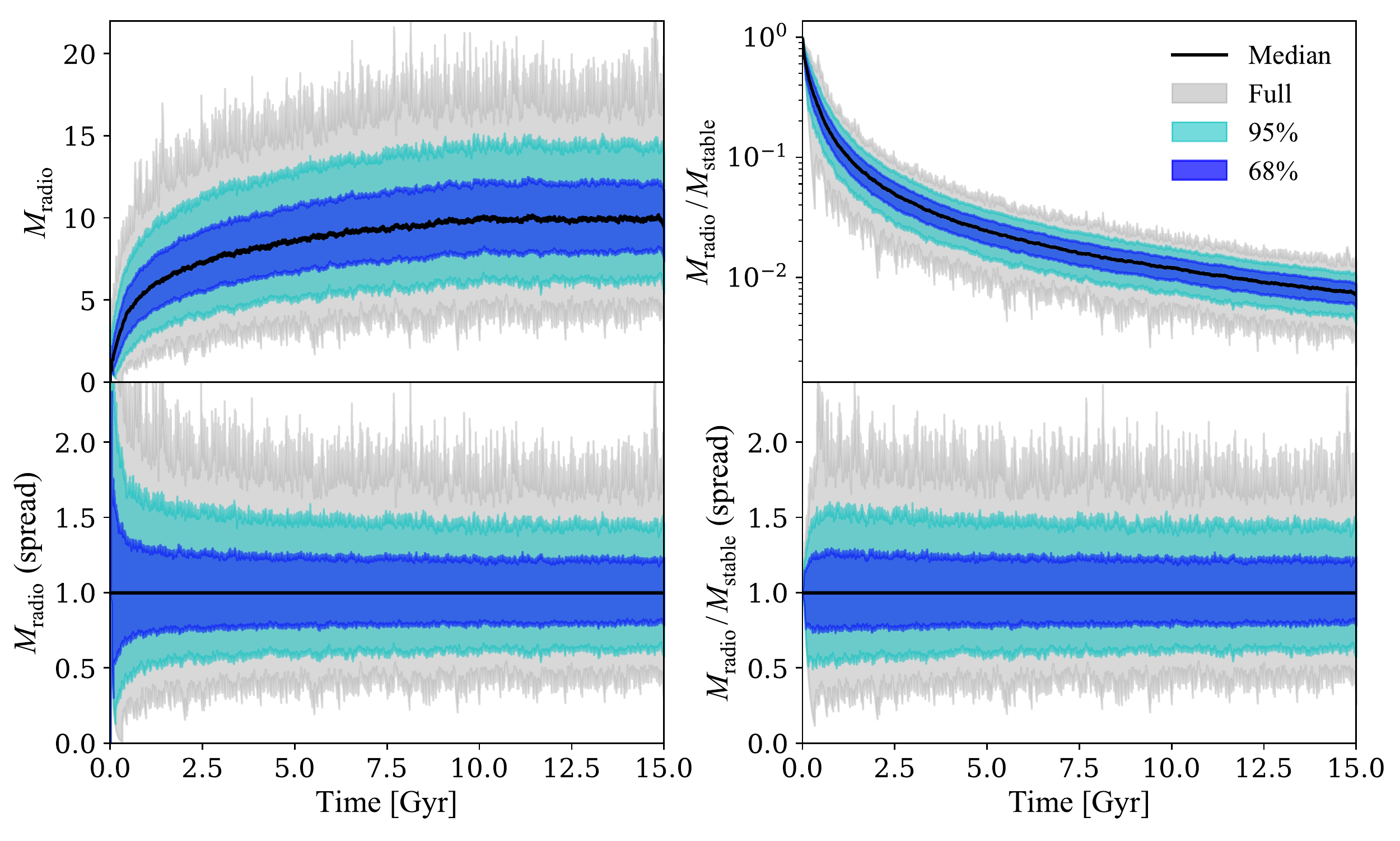}
\caption{Top panels: Temporal evolution of the median and the 68\,\% and 95\,\% confidence levels of the mass of radionuclei ($M_\mathrm{radio}$, top-left panel) and the mass ratio between the radionuclei and its stable reference isotope ($M_\mathrm{radio}/M_\mathrm{stable}$, top-right panel), using 1000 Monte Carlo runs with $\gamma=10$\,Myr and the 50\,Myr\,$-$\,10\,Gyr power-law DTD function. The grey shaded area represents the maximum and minimum values reached during the calculations. This represents an arbitrary case assuming yield production ratio of 1. Bottom panels: Relative spread around the median of $M_\mathrm{radio}$ (bottom-left panel) and $M_\mathrm{radio}/M_\mathrm{stable}$ (bottom-right panel), as a function of time using the same setup as for the upper panels.}
\label{fig:convergence_2}
\end{figure*}

%%%%%%%%%%%%%%%%%%%%%%%%%%%%%%%%%%%%%
%%%%%%%%%%%%%%%%%%%%%%%%%%%%%%%%%%%%%
\subsection{Convergence Study and Adopted Setup}
\label{sec:convergence}
%%%%%%%%%%%%%%%%%%%%%%%%%%%%%%%%%%%%%
%%%%%%%%%%%%%%%%%%%%%%%%%%%%%%%%%%%%%

We checked that the statistics resulting from our calculations reached convergence, meaning that the 68\,\% and 95\,\% confidence levels remain the same if we increase the number of Monte Carlo runs. The top panels of Figure~\ref{fig:convergence} show the statistics of the evolution of $M_\mathrm{radio}$ for 100, 1000, and 10000 Monte Carlo runs, using the $3-50$\,Myr box DTD. The larger the number of runs, the better defined are the median value and the frontiers between the 68\,\% and 95\,\% confidence levels. The black empty histograms in the lower panels represent the distribution of $M_\mathrm{radio}$ taken (arbitrarily) at 12\,Gyr from the respective top panels.

When the evolution of $M_\mathrm{radio}$ is in a steady state, meaning it oscillates stochastically around a certain value, all the $M_\mathrm{radio}$ distributions taken at any time can be stacked together in order to improve the statistics. This procedure assumes that any two points in time, $t_1$ and $t_2$ with $t_2 > t_1$, can be treated as independent from one another. This cannot be done for a single Monte Carlo run, as $N_\mathrm{SLR}(t_2)$ depends explicitly on $N_\mathrm{SLR}(t_1)$ (see Equation~\ref{eq:evRadio}). However, the values of $N_\mathrm{SLR}(t_2)$ and $N_\mathrm{SLR}(t_1)$ taken from different Monte Carlo runs are independent from each other. Therefore, with enough Monte Carlo runs, we can confidently treat every point in time in the steady state as independent. The grey filled histograms in Figure~\ref{fig:convergence} represent the distribution of $M_\mathrm{radio}$ after stacking all the distributions from 12\,Gyr to 14\,Gyr. This effectively increases the number of runs by a factor equal to the number of timesteps within that time frame (here 2000), at the cost of losing the temporal profile information.

Following the outcome of this convergence study, every set of Monte Carlo calculations presented throughout the next sections contains 1000 runs. Since we explore six DTD functions and seven values for $\gamma$, we ran in total 42 sets, resulting in 42\,000 individual runs of 15\,Gyr. This series of calculations provide the list of times at which enrichment events are occurring, on top of which the abundances of radionuclei can be followed using Equation~(\ref{eq:evRadio}) with different mean-life values. 
When quantifying the spread and the distribution of $M_\mathrm{radio}$, for a given $\gamma$, $\tau$, and DTD function, we stack together all timesteps between 12 and 14\,Gyr to improve the statistics. We did not extend the stacking operation below 12\,Gyr because for some DTD functions, $M_\mathrm{radio}$ reaches a steady state after 10\,Gyr.

The top-left panel of Figure~\ref{fig:convergence_2} shows the evolution of $M_\mathrm{radio}$ for 1000 runs, using the 50\,Myr\,$-$\,10\,Gyr power-law DTD with a mean-life of 100\,Myr. Because the delay time of each enrichment event here can be as long as 10\,Gyr, the evolution of $M_\mathrm{radio}$ only reaches a steady state after $\sim$\,10\,Gyr, as opposed to  $\sim$\,50\,Myr in the case of a $3-50$\,Myr DTD function. In general, the system needs $n>5\tau/\langle\delta\rangle$ events, or a time $T = n\,\langle\delta\rangle > 5\,\tau$, to reach the steady state (see Section~\ref{sec:intro_sub}). Therefore, for the longest mean-lives, the time needed for $M_\mathrm{radio}$ to reach a steady state can be longer than the time frame spanned by the DTD function. For example, the system needs $\sim12$\,Gyr to reach a steady state with the 50\,Myr\,$-$\,10\,Gyr DTD functions when $\tau=316$\,Myr.

Throughout this work, we only present results for cases where $M_\mathrm{radio}$ either reached a steady state by the time the early Solar System forms, after $\sim$\,$8-9$\,Gyr of Galactic evolution, or was close to a steady state by that time. As an example of the latter case, shown in the bottom-left panel of Figure~\ref{fig:convergence_2} for a 10\,Gyr-long DTD function, the relative spread around the median already converges by $\sim$\,5\,Gyr, even if $M_\mathrm{radio}$ does not reach a steady state before 10\,Gyr. Such cases are therefore included in our results.

\begin{figure*}
%\center
\includegraphics[width=7.0in]{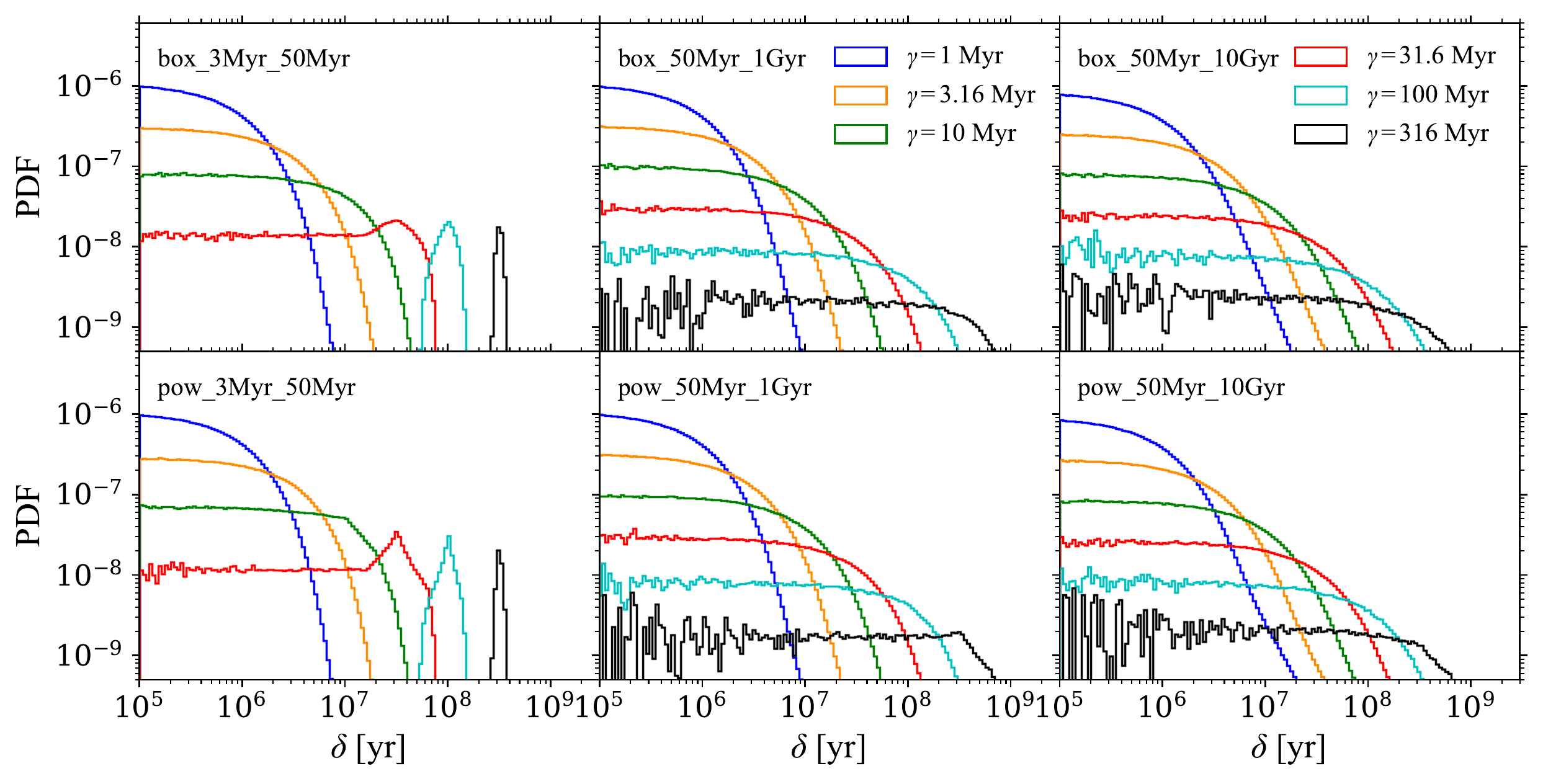}
\caption{Probability distribution function (PDF) of $\delta$, the time interval between two consecutive enrichment events (see Figure~\ref{fig:param}), as a function of the DTD function (different panels) and $\gamma$ (different colors). Each PDF was extracted by stacking the  $\delta$ distributions of 1000 Monte Carlo runs.}
\label{fig:delta}
\end{figure*}

Finally, since the present study targets isotopes that are relevant for the early Solar System, we also test our methodology with isotopic ratios in order to connect with meteorite data. The top-right panel of Figure~\ref{fig:convergence_2} shows the temporal evolution of $M_\mathrm{radio}/M_\mathrm{stable}$, the ratio between a radioactive isotope and its stable reference isotope. The decreasing trend is mainly due to the fact that the mass of stable isotopes continuously increases over time as opposed to the radioactive part which tends toward an equilibrium value (see e.g., \citealt{1984ApJ...285..411C}, \citealt{huss09}, and \citealt{cote19} for details). As shown in the bottom-right panel of Figure~\ref{fig:convergence_2}, the relative spread of $M_\mathrm{radio}/M_\mathrm{stable}$ after $\sim$\,5\,Gyr is the same as the one for $M_\mathrm{radio}$. This is because $M_\mathrm{stable}$ is insensitive to the stochastic nature of the enrichment events. Indeed, the total mass of stable isotopes after $n$ events is always the same for any run, regardless of $\gamma$ and the adopted DTD function.

%%%%%%%%%%%%%%%%%%%%%%%%%%%%%%%%%%%%%%%%
%%%%%%%%%%%%%%%%%%%%%%%%%%%%%%%%%%%%%%%%
%%%%%%%%%%%%%%%%%%%%%%%%%%%%%%%%%%%%%%%%
\section{Probability Distribution of $\delta$}
\label{sec:delta_distribution}
%%%%%%%%%%%%%%%%%%%%%%%%%%%%%%%%%%%%%%%%
%%%%%%%%%%%%%%%%%%%%%%%%%%%%%%%%%%%%%%%%
%%%%%%%%%%%%%%%%%%%%%%%%%%%%%%%%%%%%%%%%
Figure~\ref{fig:delta} shows the distribution of $\delta$ resulting from our Monte Carlo calculations as a function of $\gamma$, for the six DTD functions explored in this study. To better understand this figure, let us first define the time frame of our DTD functions as
\begin{equation}
    \Delta_\mathrm{DTD}=t_\mathrm{delay}^\mathrm{max}-t_\mathrm{delay}^\mathrm{min},
\label{eq:dtd_timeframe}
\end{equation}
where $t_\mathrm{delay}^\mathrm{max}$ and $t_\mathrm{delay}^\mathrm{min}$ are the maximum and minimum delay times of the function. This time frame allows us to predict both the maximum and minimum value of $\delta$, which includes the possibility of having simultaneous events. To understand how, we take the simple approach of having only two progenitors formed at times $t=0$ and $t=\gamma$, respectively. These progenitors will enrich the medium at times $t_1$ and $\gamma + t_2$, with $t_1$ and $t_2$ sampled from the DTD.

With this setup, the maximum possible $\delta$ is given by taking the smallest $t_1$ and the largest $t_2$ possible, and can be calculated as
\begin{equation}
    \delta_\mathrm{max}=t_\mathrm{delay}^\mathrm{max}+\gamma-t_\mathrm{delay}^\mathrm{min}=\gamma+\Delta_\mathrm{DTD}.
    \label{eq:delta_max}
\end{equation}
For the minimum possible $\delta$ we have two different cases. When $\Delta_\mathrm{DTD} < \gamma$, which means that $t_1$ is always lower than $\gamma + t_2$, $\delta_\mathrm{min}$ is given by
\begin{equation}
    \delta_\mathrm{min}=t_\mathrm{delay}^\mathrm{min}+\gamma-t_\mathrm{delay}^\mathrm{max}=\gamma-\Delta_\mathrm{DTD}.
    \label{eq:delta_min_first}
\end{equation}
When $\Delta_\mathrm{DTD} \geq \gamma$, instead, it is possible that $t_1 \geq \gamma + t_2$. But from the definition of $\delta$ (Figure~\ref{fig:param}), its value cannot be negative. Therefore, in this case,  $\delta_\mathrm{min} = 0$, meaning that two (or even more) enrichment events can occur at the same time. Both cases are summarized by the expression
\begin{equation}
    \delta_\mathrm{min}=\max(0, \gamma-\Delta_\mathrm{DTD}).
    \label{eq:delta_min}
\end{equation}
As shown in Figure~\ref{fig:delta}, all $\delta$ distributions extend down to $\delta=0$ when $\gamma\leq31.6$\,Myr, since in this case all of our DTD functions have a time frame $\Delta_\mathrm{DTD}$ longer or equal to $\sim$\,50\,Myr.  When $\gamma\geq100$\,Myr, only the longest 50\,Myr\,$-$\,1\,Gyr and 50\,Myr\,$-$\,10\,Gyr DTD functions show $\delta$ distributions extending down to $\delta=0$. Another feature of these distributions is that their width becomes larger with increasing $\gamma$, as expected from Equation~(\ref{eq:delta_max}).

\begin{deluxetable*}{lcccccc}
\tablewidth{0pc}
\tablecaption{Average time interval $\langle\delta\rangle$ between two enrichment events extracted from our Monte Carlo calculations as a function of $\gamma$ and the DTD function. Values in parenthesis are the standard deviations defined as $\sigma_\delta = \sqrt{\langle \delta^2 \rangle - \langle \delta \rangle^2}$. The full distributions of $\delta$ are shown in Figure~\ref{fig:delta}.\label{tab:delta}}
\tablehead{
\multirow{2}{*}{DTD} & \multicolumn{6}{c}{$\gamma$ [Myr]}\\
 & \colhead{1.00} & \colhead{3.16} & \colhead{10.0} & \colhead{31.6} & \colhead{100} & \colhead{316}
 }
\startdata
pow\,\_\,3Myr\,\_\,50Myr & 1.00 (0.965) & 3.16 (2.86) & 10.0 (7.76) & 31.6 (16.5) & 100 (18.1) & 316 (18.2)  \\
box\,\_\,3Myr\,\_\,50Myr & 1.00 (0.979) & 3.16 (2.96) & 10.0 (8.29) & 31.7 (17.9) & 100 (19.2) & 316 (19.2)  \\
pow\,\_\,50Myr\,\_\,1Gyr & 1.00 (0.998) & 3.16 (3.14) & 10.0 (9.78) & 31.6 (29.9) & 100 (85.4) & 315 (216)  \\
box\,\_\,50Myr\,\_\,1Gyr & 1.00 (0.999) & 3.16 (3.15) & 9.99 (9.92) & 31.6 (30.6) & 100 (91.2) & 318 (235)  \\
pow\,\_\,50Myr\,\_\,10Gyr & 1.00 (1.00) & 3.16 (3.14) & 9.98 (9.97) & 31.7 (31.1) & 99.1 (94.5) & 331 (340)  \\
box\,\_\,50Myr\,\_\,10Gyr & 1.00 (1.00) & 3.16 (3.16) & 10.0 (9.98) & 31.8 (31.7) & 101 (99.9) & 344 (349)  \\
\enddata
\end{deluxetable*}

When $\gamma\geq100$\,Myr, for the two short $3-50$\,Myr DTD functions ($\Delta_\mathrm{DTD}=47$\,Myr), $\delta_\mathrm{min}$ is greater than zero (see Equation~\ref{eq:delta_min}) and the $\delta$ distribution does not extend to $\delta=0$ (light blue lines and black lines in the left panels of Figure~\ref{fig:delta}). This means it is impossible to have two events occurring simultaneously. The maximum width of those $\delta$ distributions is then given by $\delta_\mathrm{max}-\delta_\mathrm{min}=2\Delta_\mathrm{DTD}$. As shown in Figure~\ref{fig:delta}, the peak of such distributions is more pronounced with a power-law DTD function in the form of $t^{-1}$ than with a box DTD function. This is because a delay time randomly sampled from a power law is always biased towards a certain value, as opposed to sampling a box where every delay time has the same probability. 
%%%%%%In other words, the $\delta$ distribution generated with a box DTD function will always be more \textit{uncertain} than with a power law DTD function, with both functions covering the same time frame.

Table~\ref{tab:delta} shows the quantitative values of the $\delta$ distributions shown in Figure~\ref{fig:delta}. A key feature is that, overall, $\langle\delta\rangle\approx\gamma$. In fact, for $\gamma\leq100$\,Myr, $\langle\delta\rangle=\gamma$ within 3\,\% for all DTD functions. For $\gamma=316$\,Myr, the average of $\delta$ for the longest DTD functions only deviates from $\gamma$ by about 10\,\%. In that table, $\sigma_\delta$ represents the standard deviation of $\delta$ distributions, a quantity used and defined in the analytical development presented in Section~\ref{sec:tau_gamma_grt_5}.

%%%%%%%%%%%%%%%%%%%%%%%%%%%%%%%%%%%%%%%%
%%%%%%%%%%%%%%%%%%%%%%%%%%%%%%%%%%%%%%%%
%%%%%%%%%%%%%%%%%%%%%%%%%%%%%%%%%%%%%%%%
\section{Spread of Radionuclei Under Different Enrichment Regimes}
\label{sec:spread}
%%%%%%%%%%%%%%%%%%%%%%%%%%%%%%%%%%%%%%%%
%%%%%%%%%%%%%%%%%%%%%%%%%%%%%%%%%%%%%%%%
%%%%%%%%%%%%%%%%%%%%%%%%%%%%%%%%%%%%%%%%
Here we focus on the spread of $M_\mathrm{radio}$ generated by stochastic enrichment events as a function of $\gamma$, $\tau$, and the DTD function (see Table~\ref{tab:def} for a definition of the terms used in this paper). 
In the subsections below, we classify the evolution of $M_\mathrm{radio}$ into three main regimes based on $\tau/\gamma$ (see Figure~\ref{fig:regime}) and describe their different features.  Because the results are almost identical to each other when using the box or the power-law DTD functions, we only focus on the results obtained with the box DTD functions.

Figure~\ref{fig:pdf_all} shows the $M_\mathrm{radio}$ distributions for selected values of $\tau/\gamma$. When possible, the spreads are quantified in Table~\ref{tab:uncertainties_box}, where empty cells correspond to cases of extremely long-lived isotopes: $\tau > 3$ Gyr for the two shortest DTD functions and $\tau > 1$ Gyr for the longest DTD function. For such cases, $\langle M_{\text{radio}}\rangle$ is not near a steady state by the time the Sun forms and cannot be analyzed with our statistical framework (see Section~\ref{sec:convergence}).

As a general note, any $M_\mathrm{radio}$ distribution for a given $\tau/\gamma$ is independent of $\gamma$ if the time frame $\Delta_\mathrm{DTD}$ (see Equation~\ref{eq:dtd_timeframe}) of the DTD function is significantly larger than $\gamma$. In such cases, an enrichment event can occur at any time and the delay between two consecutive events becomes almost purely random (right panels of Figure~\ref{fig:pdf_all}).
On the other hand, when the time frame of the DTD function is similar to or shorter than $\gamma$, the $M_\mathrm{radio}$ distributions is dependent on $\gamma$, and the delay between two consecutive events becomes more predictable. Indeed, if $\gamma$ is large relative to $\Delta_\mathrm{DTD}$, two enrichment events cannot occur at the same time. But if $\gamma$ becomes similar to $\Delta_\mathrm{DTD}$, there will be a probability for two or more events to occur at the same time. 

%%%%%%%%%%%%%%%%%%%%%%%%%%%%%%%%%%%%%%%%
%%%%%%%%%%%%%%%%%%%%%%%%%%%%%%%%%%%%%%%%
\subsection{Regime I: $\tau/\gamma\gtrsim2$}
\label{sec:tau_gamma_grt_5}
%%%%%%%%%%%%%%%%%%%%%%%%%%%%%%%%%%%%%%%%
%%%%%%%%%%%%%%%%%%%%%%%%%%%%%%%%%%%%%%%%

\begin{figure}
%\center
\includegraphics[width=3.4in]{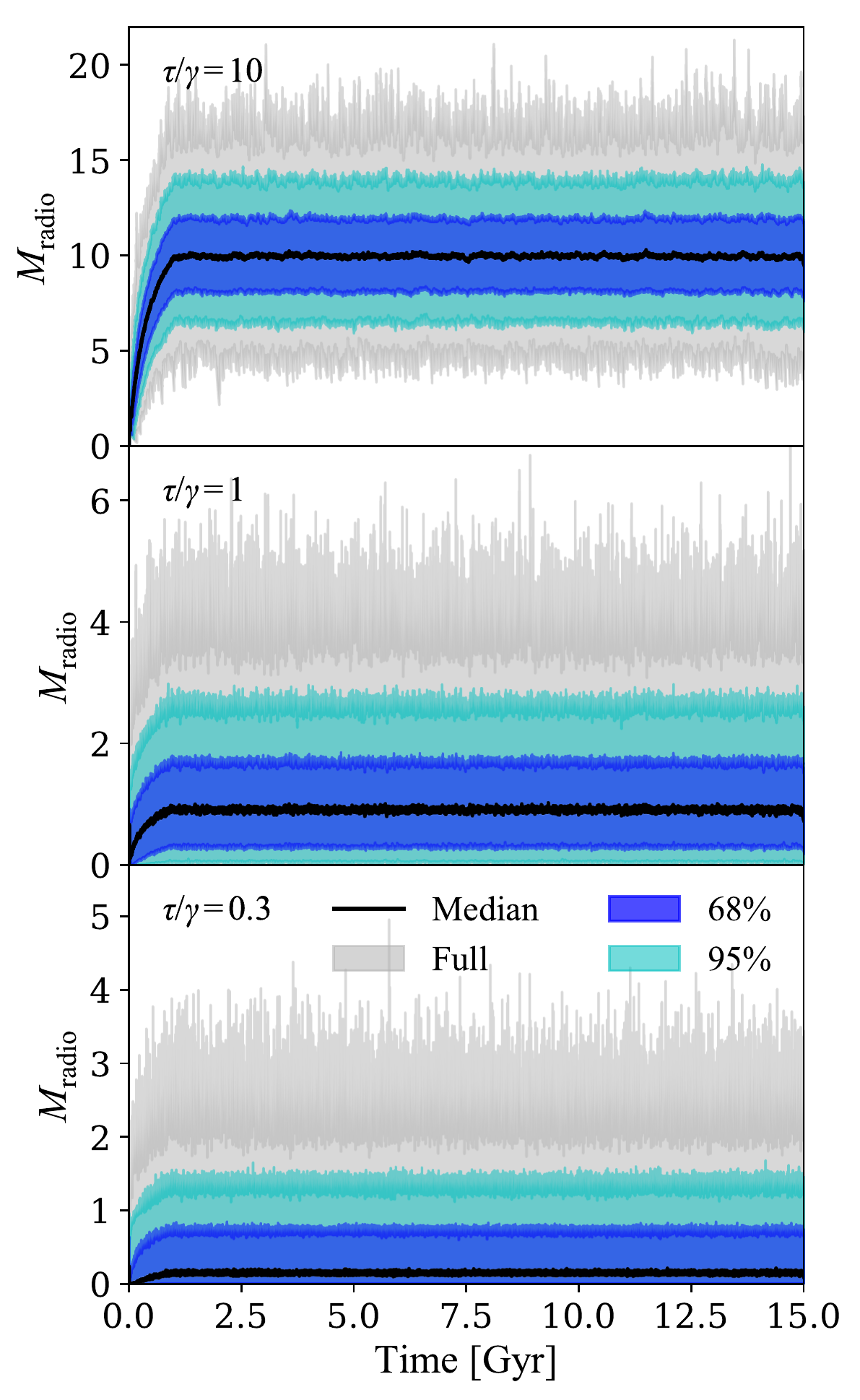}
\caption{Evolution of the median and the 68\,\% and 95\,\% confidence levels of the mass of radionuclei ($M_\mathrm{radio}$) as a function of time, using 1000 Monte Carlo runs with the 50\,Myr\,$-$\,1\,Gyr power-law DTD function, and assuming $\tau/\gamma=10$ (top panel), 1 (middle panel), and 0.3 (bottom panel). The grey shaded area represents the maximum and minimum values reached during the calculations.}
\label{fig:regime}
\end{figure}

\begin{figure*}
%\center
\includegraphics[width=7.0in]{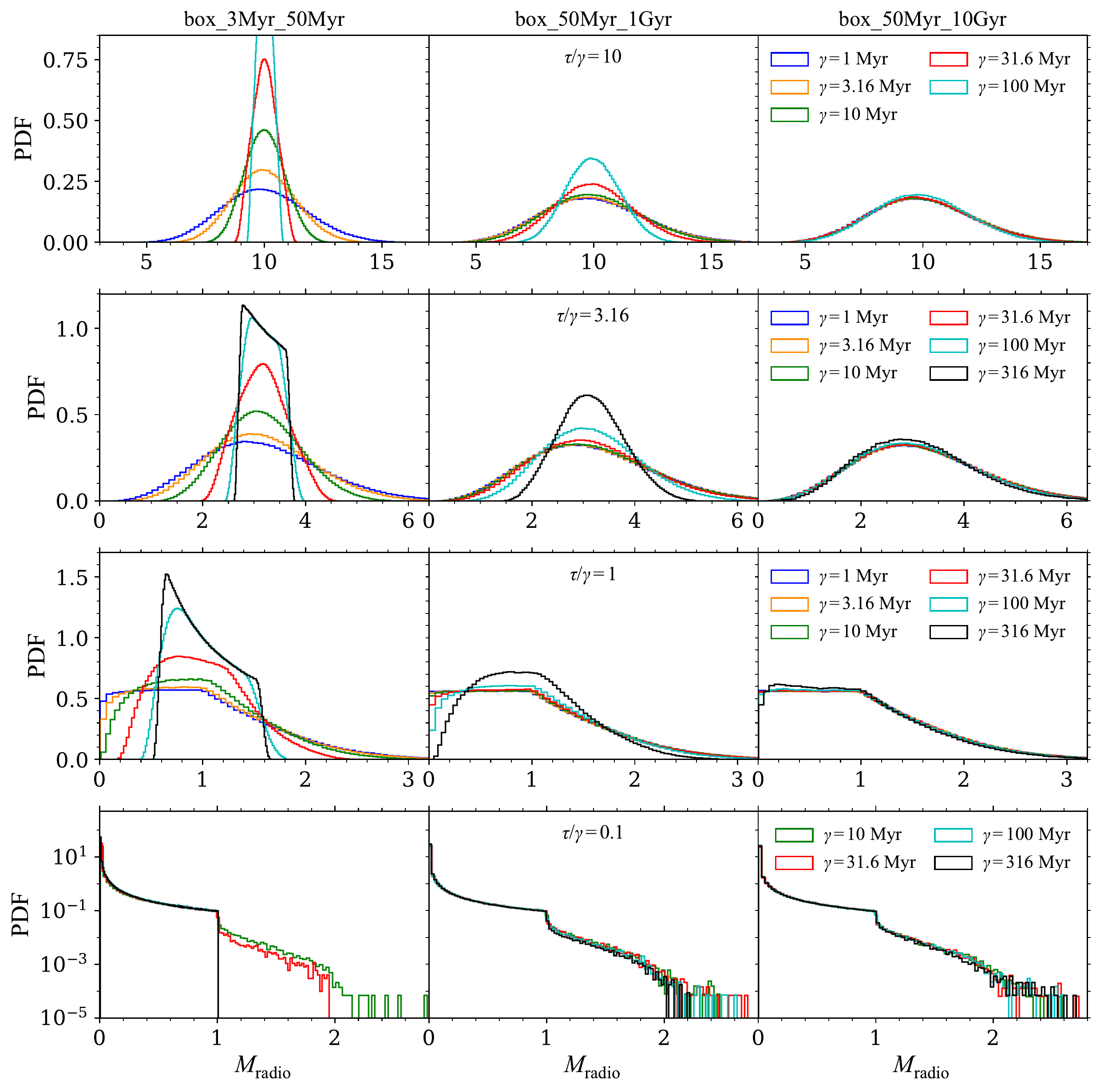}
\caption{Probability distribution function (PDF) of $M_\mathrm{radio}$ as a function of the DTD box function (different columns), $\tau/\gamma$ (different rows), and $\gamma$ (different colors). From left to right, the column titles represent the short $3-50$\,Myr, the medium 50\,Myr\,$-$\,1\,Gyr, and the long 50\,Myr\,$-$\,10\,Gyr DTD functions. Each PDF (each line) represents the distribution derived from 1000 Monte Carlo runs (see Section~\ref{sec:convergence}). Each row does not show the same number of $\gamma$ values because of the defined range for $\gamma$ and $\tau$ (see Section~\ref{sec:dtd_meanlives}). The PDFs using the power-law DTD functions are always almost identical to those shown in this figure using the box DTD functions.}
\label{fig:pdf_all}
\end{figure*}

If $\tau/\gamma\gtrsim 2$, it is possible to define the statistics of the distributions as a function of the DTD function (two upper rows in Figure~\ref{fig:regime}). The corresponding spreads are reported in Table~\ref{tab:uncertainties_box} (see also Section~\ref{sec:comparison}). This regime can also be applied in some cases for $\tau/\gamma\,=1$ when using the two shortest DTD functions. The change at $\tau/\gamma\,=1$ in the table from reporting spreads to reporting upper limits marks the transition between Regime I and Regime II, which is discussed below.

When $\tau/\gamma\gtrsim5$, the distribution of $M_\mathrm{radio}$ is symmetric around the steady-state value (top panel of Figure~\ref{fig:regime}), which is given by $\tau/\langle \delta \rangle\approx\tau/\gamma$ (see Section~\ref{sec:delta_distribution}). In this regime, the steady-state value is large enough that the minimum $M_\mathrm{radio}$ values always remain above zero. In other words, the SLR never have time to completely decay before the next enrichment event. In this case, we can derive an analytical solution for the distribution, to be compared to the numerical results.

First, we recover a fairly similar expression for the average value to that obtained from Equation~(\ref{eq:constDeltModel}), but for the case when $\delta$ is a random variable. We find that, in the steady state\footnote{The derivation of $\langle N_\mathrm{SLR} \rangle$, $\langle N_\mathrm{SLR}^2 \rangle$, and $\sigma$ can be found in Appendix \ref{sec:longMaths}.},
\begin{equation}
    \langle N_\mathrm{SLR} \rangle \approx \frac{\tau}{\langle \delta \rangle}.
    \label{eq:avgNSLR}
\end{equation}

\noindent On the other hand, the standard deviation $\sigma = \sqrt{\langle N_\mathrm{SLR}^2\rangle - \langle N_\mathrm{SLR}\rangle^2}$, which is a measure of the uncertainty, differs substantially from that of the simple model with constant $\delta_c$. The full expression in the steady state for $\langle N_\mathrm{SLR}^2\rangle$ is
\begin{equation}
    \langle N_\mathrm{SLR}^2 \rangle \approx \frac{\tau}{2\langle \delta \rangle}\left(1 + 2\frac{\langle e^{-\delta/\tau}\rangle}{1 - \langle e^{-\delta/\tau}\rangle}\right).
\end{equation}
With this expression, we can write $\sigma$ as
\begin{equation}
    \sigma = \sqrt{\frac{\tau}{2\langle \delta \rangle}}\sqrt{\frac{1 + \langle e^{-\delta/\tau} \rangle}{1 - \langle e^{-\delta/\tau} \rangle} - \frac{2\tau}{\langle \delta \rangle}},
\end{equation}
which, in the case of $\tau/\langle \delta \rangle \gtrsim 3$,
%(i.e., $\langle \delta \rangle/\tau$ is small enough to approach the exponential to the second-order Taylor expansion) 
can be approximated to
\begin{equation}
    \sigma \approx \frac{\sigma_\delta}{\langle \delta \rangle} \sqrt{\frac{\tau}{2\langle \delta \rangle}},
    \label{eq:absSpread}
\end{equation}
where $\sigma_\delta = \sqrt{\langle \delta^2 \rangle - \langle \delta \rangle^2}$. According to our Monte Carlo calculations, $\sigma_\delta \sim \langle \delta \rangle$ for most cases (see Table \ref{tab:delta}), which allows $\sigma$ to be expressed independently of $\sigma_\delta$ as
\begin{equation}
    \sigma \approx \sqrt{\frac{\tau}{2\langle \delta \rangle}}.
\end{equation}
For bell-shaped distributions, we can approximate the uncertainty around $\langle N_\mathrm{SLR} \rangle$ with $\sigma$. In this case, we have that the relative uncertainty of
\begin{equation}
    \frac{\sigma}{\langle N_\mathrm{SLR} \rangle} \approx \frac{\sigma_\delta}{\langle \delta \rangle} \sqrt{\frac{\langle \delta \rangle}{2\tau}}.
    \label{eq:relSpread}
\end{equation}

In summary, there are three conditions for using Equation~(\ref{eq:relSpread}) as a proxy for the relative uncertainty: (1) the steady-state regime is reached; (2) $\tau/\langle \delta \rangle$ is larger than $\sim5$; and (3) $\sigma$ is a good predictor of the uncertainty, which as shown in Section~\ref{sec:delta_distribution} is true except for the short $3-50$\,Myr DTD functions when $\gamma$ is greater than the time frame $\Delta_\mathrm{DTD}$. The latter case corresponds to the narrow $\delta$ distributions that do not extend down to $\delta=0$ in Figure~\ref{fig:delta}.

As derived in Equation~(\ref{eq:relSpread}), the relative spread of $M_\mathrm{radio}$ depends on both $\sqrt{\gamma/2\tau}$ and $\sigma_\delta/\gamma$, where $\sigma_\delta$ is the standard deviation of the $\delta$ distribution. It follows that in the $\tau/\gamma\gtrsim5$ regime, when using long DTD functions, the width of the $M_\mathrm{radio}$ distribution does not depend on $\gamma$. This is because $\sigma_\delta$ is always similar to $\gamma$ in these cases (see Table~\ref{tab:delta}). The relative spread is thus only given by $\sqrt{\gamma/2\tau}$, since $\sigma_\delta/\gamma\approx1$. However, in the case of the short $3-50$\,Myr DTD function, $\sigma_\delta/\gamma<1$, which generates narrower spreads compared to longer DTD functions. Furthermore, because $\sigma_\delta/\gamma$ decreases when $\gamma$ increases, the $M_\mathrm{radio}$ spreads depend in this case on $\gamma$. This major difference between short and long DTD functions can be traced back to their different $\delta$ distributions (see Figure~\ref{fig:delta} and explanations in Section~\ref{sec:delta_distribution}). 

%%%%%%%%%%%%%%%%%%%%%%%%%%%%%%%%%%%%%%%%%%%%
%%%%%%%%%%%%%%%%%%%%%%%%%%%%%%%%%%%%%%%%%%%%
\subsection{Abundance Spread Quantification for Regime I}
\label{sec:comparison}
\begin{figure}
%\center
\includegraphics[width=3.37in]{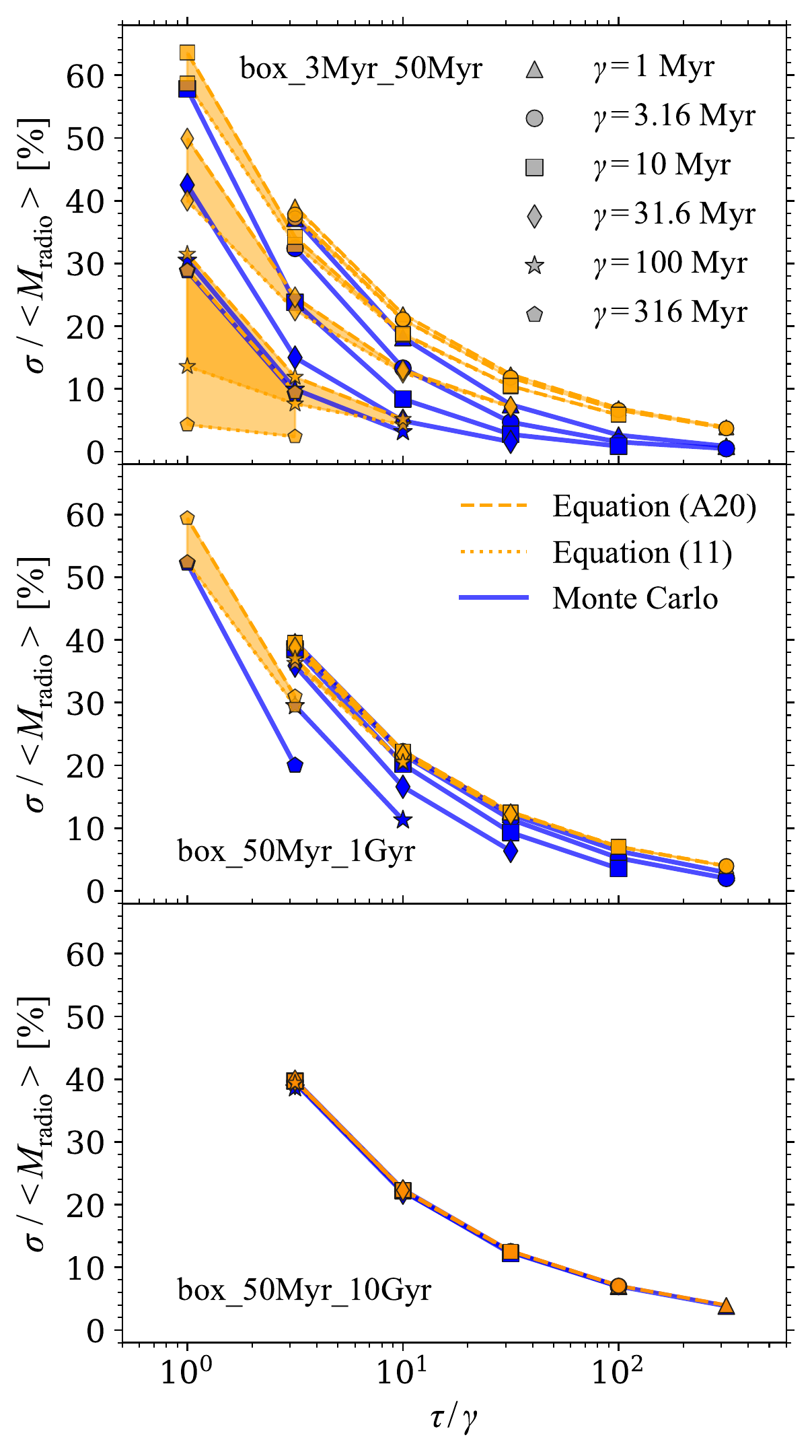}
\caption{Relative standard deviation (in percentage) of $M_\mathrm{radio}$ caused by stochastic enrichment events, as a function $\tau/\gamma$ for different $\gamma$ (different symbols) and different box DTD functions (different panel). Here, $\langle M_\mathrm{radio} \rangle$ is the average equilibrium value of $M_\mathrm{radio}$, while $\sigma$ is its standard deviation. We only show cases where the $M_\mathrm{radio}$ distribution does not extend down to zero. For the orange dashed and dotted lines, $\sigma$ was calculated with Equations~\ref{eq:sqAvgProof2} and \ref{eq:absSpread}, respectively, using the $\delta$ distributions of the Monte Carlo calculations. For the blue lines, $\sigma$ was calculated from the following definition, $\sqrt{\langle M_\mathrm{radio}^2\rangle-\langle M_\mathrm{radio}\rangle^2}$, using the $M_\mathrm{radio}$ distribution of the Monte Carlo calculations. The orange bands are eye guides to help visualizing the impact of using Equations~\ref{eq:sqAvgProof2} and \ref{eq:absSpread} for the same $\gamma$. We note that those two equations should be used only when $\tau/\gamma\gtrsim3$, as otherwise the spread around the median is non-symmetric, which means the uncertainty should be quantified by the non-symmetric 68\,\% confidence level, rather than by the standard deviation provided by these equations.}
\label{fig:mc_vs_full}
\end{figure}

%%%%%%%%%%%%%%%%%%%%%%%%%%%%%%%%%%%%%%%%%%%%
%%%%%%%%%%%%%%%%%%%%%%%%%%%%%%%%%%%%%%%%%%%%
Figure~\ref{fig:mc_vs_full} shows the relative standard deviation of $M_\mathrm{radio}$ for $\tau/\gamma>1$, where the spread is defined by the standard deviation of the $M_\mathrm{radio}$ distributions. For $\tau/\gamma\gtrsim5$, the spread is symmetric and the standard deviations are similar to the 68\,\% confidence levels presented in Table~\ref{tab:uncertainties_box}. However, for $\tau/\gamma$ below 5, the spread becomes non-symmetric and we refer to Table~\ref{tab:uncertainties_box} rather than Figure~\ref{fig:mc_vs_full} for a more precise quantification. This figure also compares the standard deviations obtained from the $M_\mathrm{radio}$ distributions of the Monte Carlo calculations with the analytical solutions derived in Section~\ref{sec:tau_gamma_grt_5}, which only require as an input the $\delta$ distributions extracted from our Monte Carlo calculations. As expected from Figure~\ref{fig:pdf_all}, the standard deviation only depends on $\gamma$ in the case of the short DTD function.

Equations~(\ref{eq:avgNSLR}) and (\ref{eq:relSpread}) are calculated as the temporal average and standard deviation for a single SLR evolutionary history track. In contrast, the Monte Carlo simulations calculate the average and spread (i.e., 68\,\% confidence level) of several independent tracks in a single moment in time. As seen from Figure~\ref{fig:mc_vs_full}, these two approaches are essentially identical.
In the figure we show the analytical results computed using both our complete analytical solution (Equation~\ref{eq:sqAvgProof2} derived in the Appendix), and the simplified approximated solution (Equations~\ref{eq:absSpread} and \ref{eq:relSpread}). For long DTD functions (lower panel of Figure~\ref{fig:mc_vs_full}), both analytical solutions are consistent with the spread derived from the Monte Carlo calculations, while for short DTD functions (upper panel), the analytical solutions start to differ from the Monte Carlo calculations and the complete analytical solution (Equation~\ref{eq:sqAvgProof2}) always provides an upper limit for the uncertainty. 

Overall, Figure~\ref{fig:mc_vs_full} shows that when $\tau/\gamma>1$, and when the spread can be quantified by confidence intervals (see Table~\ref{tab:uncertainties_box}), the uncertainty in the abundances of radionuclei caused by stochasticity is always below $\sim$\,60\%. For $\tau/\gamma>10$, the uncertainty drops below 20\,\%. This is assuming a constant star formation history as well as enrichment events that always add the same amount of radioactive material in the given parcel of interstellar matter.

%%%%%%%%%%%%%%%%%%%%%%%%%%%%%%%%%%%%%%%%%%%%%%%%%%%%%%%%%
%%%%%%%%%%%%%%%%%%%%%%%%%%%%%%%%%%%%%%%%%%%%%%%%%%%%%%%%%
\subsection{Regime II: $0.3\lesssim\tau/\gamma\lesssim2$}
\label{sec:tau_gamma_ll1}
%%%%%%%%%%%%%%%%%%%%%%%%%%%%%%%%%%%%%%%%%%%%%%%%%%%%%%%%%
%%%%%%%%%%%%%%%%%%%%%%%%%%%%%%%%%%%%%%%%%%%%%%%%%%%%%%%%%

In this regime, the $M_\mathrm{radio}$ distribution is never symmetric (e.g., middle panel of Figure~\ref{fig:regime}). While the distribution above the steady-state value can extend upward, the distribution below that value is now limited to zero. In other words, the SLR can have the time to \textit{completely} decay before the next enrichment event.
In fact, as shown in the third row of Figure~\ref{fig:pdf_all} ($\tau/\gamma=1$), several of the $M_\mathrm{radio}$ distributions extend down to zero. As for the symmetric regime described above, the distributions calculated using the longest DTD function do not depend on $\gamma$ for a given $\tau/\gamma$ value, while those using the shortest DTD function do. Furthermore, for the shortest DTD functions, and for the longest $\gamma$ values, $\sigma_\delta/\gamma$ becomes so small that the width of the $M_\mathrm{radio}$ distribution does not extend down to zero. Those cases belong to Regime I and are shown in Figure~\ref{fig:mc_vs_full}.

For Regime II, a different approach is required to define the probability distribution of $M_\mathrm{radio}$, since the abundance carry the signature of only a few events that need to be combined statistically. This approach should be applied to specific cases rather than general cases. We refer to C\^ot\'e et al. (in preparation) for a first application with $r$-process isotopes. We plan to apply this case to $s$-process isotopes in future work.

%%%%%%%%%%%%%%%%%%%%%%%%%%%%%%%%%%%%%%%%%%%%%%%%%%%%%%
%%%%%%%%%%%%%%%%%%%%%%%%%%%%%%%%%%%%%%%%%%%%%%%%%%%%%%
\subsection{Regime III: $\tau/\gamma\lesssim0.3$}
\label{sec:tau_gamma_ll1}
%%%%%%%%%%%%%%%%%%%%%%%%%%%%%%%%%%%%%%%%%%%%%%%%%%%%%%
%%%%%%%%%%%%%%%%%%%%%%%%%%%%%%%%%%%%%%%%%%%%%%%%%%%%%%

The last regime occurs when $\tau/\gamma\lesssim0.3$ (bottom panel of Figure~\ref{fig:regime}), when the SLRs decay almost completely before the next enrichment event. As shown in the bottom row of Figure~\ref{fig:pdf_all}, all $M_\mathrm{radio}$ distributions are piled down to zero, regardless of $\gamma$ and the adopted DTD function. In addition, all distributions show a sharp drop at $M_\mathrm{radio}=1$\,M$_{\odot}$. Since the contribution of a previous event quickly becomes negligible, a new enrichment event will most likely reset $M_\mathrm{radio}$ to the mass of radionuclei ejected by one event, which is assumed to be 1 in our study\footnote{This assumption was made so that follow-up studies can re-normalize our $M_\mathrm{radio}$ distributions to any physically-motivated yields.}. It is nevertheless possible for two or three events to occur at the same time, which is shown by the low-probability tail at $M_\mathrm{radio}>1$. This tail, however, is not present with the short $3-50$\,Myr DTD function when $\gamma=100$ and 316\,Myr. This is because the time frame of this DTD is shorter than $\gamma$, which makes it impossible for two consecutive events to occur at the same time (see the $\delta$ distributions in Figure~\ref{fig:delta}, the light blue lines and black lines in the left panels). 

Overall, the median value of the $M_\mathrm{radio}$ distributions quickly diverges from $\tau/\gamma$ when this ratio drops below 1. For the most extreme cases (e.g., $\tau/\gamma=0.01$), the relative spread can be larger than a factor of $10^{20}$. In this regime, the quantification of the spread is therefore irrelevant and we do not include it in Table~\ref{tab:uncertainties_box}. The relevant statistical quantity in Regime III is instead the number of events that is typically carried by $M_\mathrm{radio}$. 

\begin{figure*}
%\center
\includegraphics[width=7.0in]{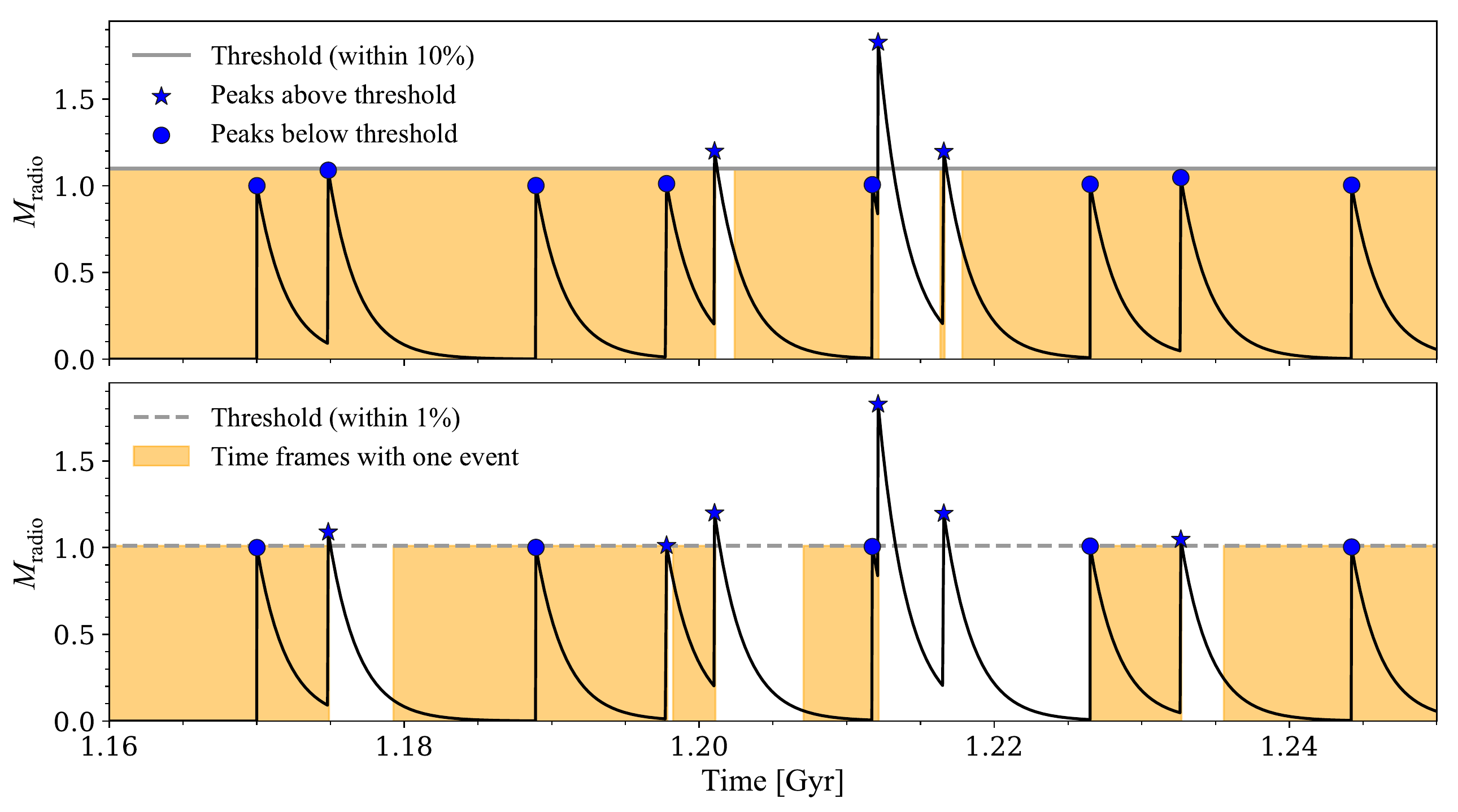}
\caption{Zoom-in of one arbitrary Monte Carlo realization for $\tau/\gamma=0.2$ to visualize our method to calculate the probabilities for $M_\mathrm{radio}$ to carry the signature of only one event (shown in Figure~\ref{fig:one_event}). The orange bands represent the time frames when $M_\mathrm{radio}$ only carries the signature of \textit{one} event, with less than 10\,\% (upper panel) or 1\,\% (lower panel) contribution coming from other previous events. The sum of all those time frames divided by the duration of the simulations gives the \textit{time fraction} probabilities (orange bands) in Figure~\ref{fig:one_event}. The blue \textit{peak fraction} probabilities in the Figure~\ref{fig:one_event} are calculated by dividing the number of peaks below $M_\mathrm{radio}=1.1$ or $M_\mathrm{radio}=1.01$ (blue dots in the current figure) by the total number of peaks (blue dots and stars). Our two approaches are applied only within the periods of time when $M_\mathrm{radio}$ is in equilibrium.}
\label{fig:one_event_explanation}
\end{figure*}

\begin{figure}
%\center
\includegraphics[width=3.37in]{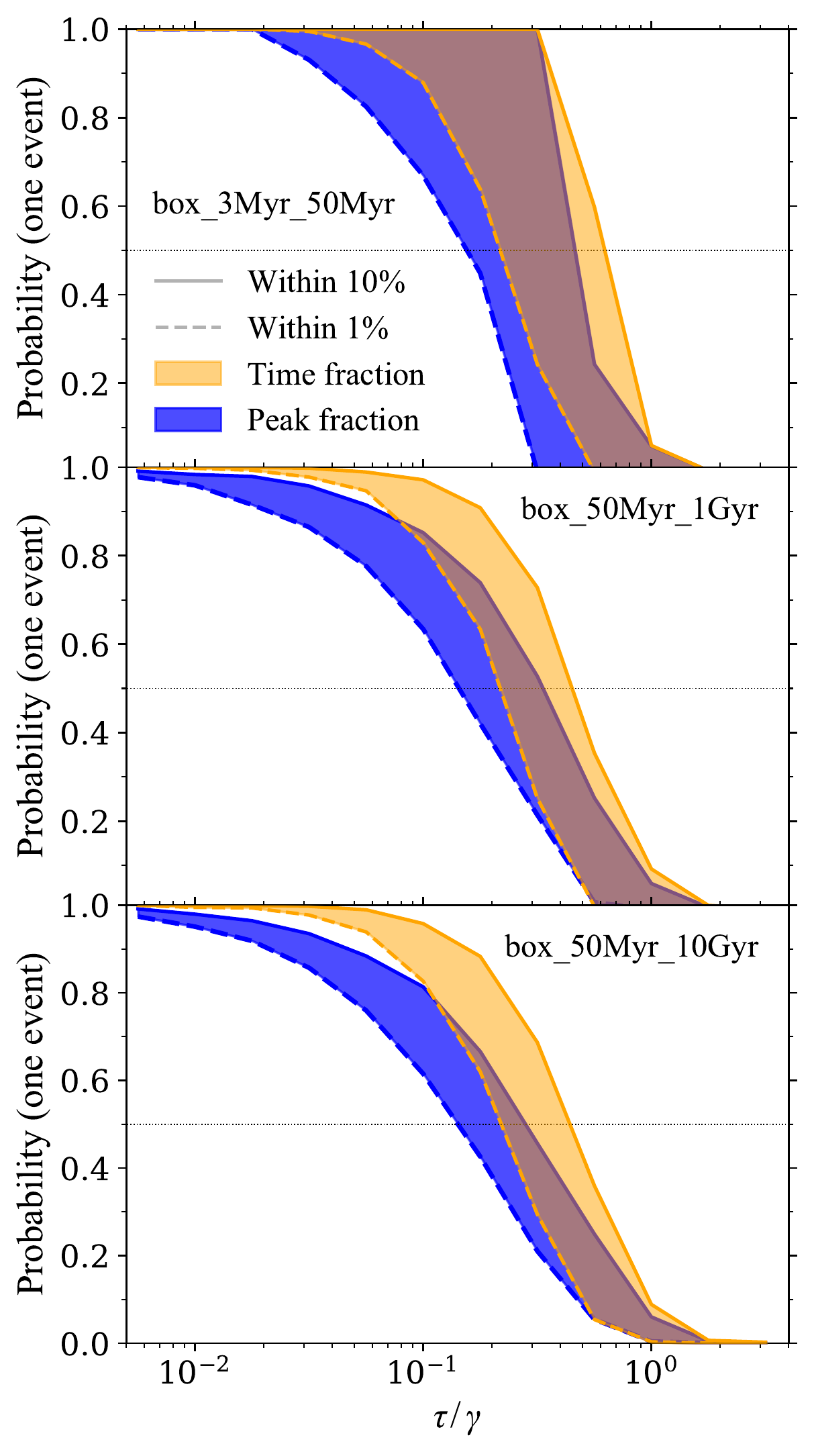}
\caption{Probability of $M_\mathrm{radio}$ to carry the signature of only one enrichment event, as a function of $\tau/\gamma$ for our three different box DTD functions (different panels). The orange bands represent the fraction of the time when $M_\mathrm{radio}$ carried only event, within 10\,\% (solid lines) and 1\,\% (dashed lines) accuracy. The blue bands represent the fraction of abundance peaks that fell below $M_\mathrm{radio}=1.1$ (solid lines) and $M_\mathrm{radio}=1.01$ (dashed lines). We refer to Figure~\ref{fig:one_event_explanation} for more details on those two approaches. The horizontal thin dotted lines mark the 50\,\% probability.}
\label{fig:one_event}
\end{figure}

To calculate this, at any time $t$, we define that $M_\mathrm{radio}$ carries the signature of one event if it is almost entirely composed of the ejecta coming from the last event (LE) that occurred before $t$. To obtain this composition, we calculate the contribution of all the events that occurred before LE, using Equation~(\ref{eq:evRadioFirst}). If the collective contribution of those events to $M_\mathrm{radio}$ at time $t$ is less than a given threshold $f_\mathrm{thresh}$, we assume that $M_\mathrm{radio}$ effectively only carries the contribution of the LE. From a list of enrichment event times generated using our Monte Carlo framework, we can quantify analytically the total amount of time when $M_\mathrm{radio}$ is only composed of \textit{one} event. By dividing this time by the total duration of our simulations, we obtain the probability for $M_\mathrm{radio}$ to carry only one event, at any given time. A visualization of our methodology is given in Figure~\ref{fig:one_event_explanation}. The orange bands in Figure~\ref{fig:one_event} show the range of probabilities when $f_\mathrm{thresh}$ is varied from 10\,\% to 1\,\%.  We note that we cannot simply assume $f_\mathrm{thresh}=0$, because each progenitor has numerically a non-zero contribution at any given time following their enrichment event. Note that these probabilities are purely theoretical as they do not account for a possible lower-limit abundance thresholds below which the detection of radionuclei would be experimentally or observationally impossible. 

An alternative approach to generate the probabilities is to scan all enrichment events and calculate the fraction of abundance peaks below $M_\mathrm{radio}=1+f_\mathrm{thresh}$ (blue dots in Figure~\ref{fig:one_event_explanation}). The disadvantage of this alternative approach, however, is that it does not account for any time frame. 

Figure~\ref{fig:one_event} shows the probability for $M_\mathrm{radio}$ to carry the signature of only one event as a function of $\tau/\gamma$ for the box DTD functions. Results are similar when using the power-law DTD functions. As shown by the blue bands in Figure~\ref{fig:one_event}, the second approach described above provides typically slightly lower probabilities. Overall, regardless of the approach used, it is clear that $M_\mathrm{radio}$ likely carries more than one event when $\tau/\gamma>1$, while there are probabilities $>$\,80\,\% that $M_\mathrm{radio}$ only carries one event when $\tau/\gamma<0.1$.

\begin{deluxetable*}{lcccccc}
\tablewidth{0pc}
\tablecaption{Monte Carlo spread for the radioactive abundance distributions presented in Figure~\ref{fig:pdf_all} as a function of $\gamma$ (different columns) and $\tau/\gamma$ (different rows) for the three box DTD functions. Cells with a plus and minus value show the median and the absolute 68\,\% confidence levels. For distributions when there is roughly equal probability to have any value between 0 and 1, we report the upper limit below which 84\,\% of the distribution is contained. Empty cells represent cases where $\tau$ is so long that the abundance is not in or near a steady state by the time of the formation of the Sun. We assumed that every enrichment event ejects 1\,$M_\odot$ of SLR. Therefore, all values presented in this table should be multiplied by the appropriate yields.\label{tab:uncertainties_box}}
\tablehead{
\multirow{2}{*}{$\tau/\gamma$} & \multicolumn{6}{c}{$\gamma$ [Myr]} \\
 & \colhead{1} & \colhead{3.16} & \colhead{10} & \colhead{31.6} & \colhead{100} & \colhead{316}
 }
\startdata
 & \multicolumn{6}{c}{box\,\_\,3Myr\,\_\,50Myr} \\[0.05cm]
316   & $316\,^{+\,2.76}_{-\,2.75}$ & $316\,^{+\,1.57}_{-\,1.57}$ &  &  & &  \\
100   & $100\,^{+\,2.66}_{-\,2.64}$ & $100\,^{+\,1.56}_{-\,1.55}$ & $100\,^{+\,0.90}_{-\,0.90}$ & & & \\
31.6  & $31.6\,^{+\,2.40}_{-\,2.34}$ & $31.6\,^{+\,1.51}_{-\,1.48}$ & $31.6\,^{+\,0.89}_{-\,0.89}$ & $31.6^{+\,0.54}_{-\,0.53}$ & &  \\
10    & $9.93\,^{+\,1.89}_{-\,1.76}$ & $9.97\,^{+\,1.37}_{-\,1.30}$ & $9.99\,^{+\,0.87}_{-\,0.84}$ & $9.99\,^{+\,0.53}_{-\,0.52}$ & $10.0\,^{+\,0.35}_{-\,0.34}$ &  \\
3.16  & $3.06\,^{+\,1.27}_{-\,1.08}$ & $3.10\,^{+\,1.10}_{-\,0.96}$ & $3.13\,^{+\,0.80}_{-\,0.73}$ & $3.15\,^{+\,0.51}_{-\,0.49}$ & $3.15\,^{+\,0.36}_{-\,0.33}$ & $3.15^{+\,0.36}_{-\,0.32}$ \\
1     & $<1.68$ & $<1.66$ & $0.93\,^{+\,0.65}_{-\,0.53}$ & $0.96\,^{+\,0.47}_{-\,0.41}$ & $0.96\,^{+\,0.39}_{-\,0.28}$ & $0.95\,^{+\,0.39}_{-\,0.27}$ \\
%1     & $0.90\,^{+\,0.78}_{-\,0.60}$ & $0.91\,^{+\,0.75}_{-\,0.58}$ & $0.93\,^{+\,0.65}_{-\,0.53}$ & $0.96\,^{+\,0.47}_{-\,0.41}$ & $0.96\,^{+\,0.39}_{-\,0.29}$ & $0.95\,^{+\,0.39}_{-\,0.27}$ \\
%0.316 & -- & $0.15\,^{+\,0.57}_{-\,0.14}$ & $0.16\,^{+\,0.55}_{-\,0.15}$ & $0.19\,^{+\,0.49}_{-\,0.16}$ & $0.22\,^{+\,0.42}_{-\,0.15}$ & $0.21\,^{+\,0.41}_{-\,0.14}$ \\
\hline
& \multicolumn{6}{c}{box\,\_\,50Myr\,\_\,1Gyr} \\[0.05cm]
316   & $316\,^{+\,9.38}_{-\,9.08}$ & $316\,^{+\,6.39}_{-\,6.29}$ &  &  &  & \\
100   & $99.9\,^{+\,6.46}_{-\,6.25}$ & $99.9\,^{+\,5.28}_{-\,5.16}$ & $100\,^{+\,3.61}_{-\,3.52}$ &  &  &  \\
31.6  & $31.5\,^{+\,3.95}_{-\,3.72}$ & $31.5\,^{+\,3.67}_{-\,3.51}$ & $31.6\,^{+\,3.02}_{-\,2.87}$ & $31.7\,^{+\,2.04}_{-\,2.01}$ &  &  \\
10    & $9.89\,^{+\,2.32}_{-\,2.10}$ & $9.89\,^{+\,2.25}_{-\,2.06}$ & $9.93\,^{+\,2.11}_{-\,1.94}$ & $9.96\,^{+\,1.72}_{-\,1.61}$ & $9.97\,^{+\,1.17}_{-\,1.12}$ &  \\
3.16  & $3.05\,^{+\,1.36}_{-\,1.14}$ & $3.05\,^{+\,1.34}_{-\,1.13}$ & $3.06\,^{+\,1.31}_{-\,1.12}$ & $3.08\,^{+\,1.22}_{-\,1.06}$ & $3.11\,^{+\,0.99}_{-\,0.89}$ & $3.13\,^{+\,0.67}_{-\,0.62}$ \\
1     & $<1.70$ & $<1.69$ & $<1.69$ & $<1.68$ & $<1.64$ & $0.94\,^{+\,0.58}_{-\,0.49}$ \\
%1     & $0.89\,^{+\,0.80}_{-\,0.60}$ & $0.89\,^{+\,0.80}_{-\,0.60}$ & $0.89\,^{+\,0.80}_{-\,0.60}$ & $0.89\,^{+\,0.78}_{-\,0.59}$ & $0.91\,^{+\,0.72}_{-\,0.57}$ & $0.91\,^{+\,0.59}_{-\,0.48}$ \\
%0.316 & -- & $0.14\,^{+\,0.58}_{-\,0.14}$ & $0.14\,^{+\,0.58}_{-\,0.14}$ & $0.14\,^{+\,0.58}_{-\,0.14}$ & $0.15\,^{+\,0.56}_{-\,0.14}$ & $0.15\,^{+\,0.53}_{-\,0.14}$ \\
\hline
& \multicolumn{6}{c}{box\,\_\,50Myr\,\_\,10Gyr} \\[0.05cm]
316   & $315\,^{+\,12.3}_{-\,11.9}$ &  &  &  &  &  \\
100   & $99.7\,^{+\,7.12}_{-\,6.85}$ & $100\,^{+\,6.95}_{-\,6.64}$ &  &  &  &  \\
31.6  & $31.4\,^{+\,4.06}_{-\,3.83}$ & $31.6\,^{+\,4.02}_{-\,3.81}$ & $31.4\,^{+\,3.97}_{-\,3.69}$ &  &  &  \\
10    & $9.87\,^{+\,2.34}_{-\,2.12}$ & $9.91\,^{+\,2.32}_{-\,2.10}$ & $9.86\,^{+\,2.34}_{-\,2.08}$ & $9.85\,^{+\,2.27}_{-\,2.06}$ &  &  \\
3.16  & $3.04\,^{+\,1.36}_{-\,1.14}$ & $3.06\,^{+\,1.36}_{-\,1.14}$ & $3.04\,^{+\,1.36}_{-\,1.13}$ & $3.04\,^{+\,1.34}_{-\,1.13}$ & $3.03\,^{+\,1.29}_{-\,1.10}$ &  \\
1     & $<1.70$ & $<1.70$ & $<1.70$ & $<1.69$ & $<1.67$ & $<1.62$ \\
%1     & $0.89\,^{+\,0.81}_{-\,0.61}$ & $0.89\,^{+\,0.80}_{-\,0.60}$ & $0.89\,^{+\,0.80}_{-\,0.60}$ & $0.89\,^{+\,0.79}_{-\,0.60}$ & $0.86\,^{+\,0.79}_{-\,0.59}$ & $0.86\,^{+\,0.77}_{-\,0.59}$ \\
%0.316 & -- & $0.14\,^{+\,0.58}_{-\,0.14}$ & $0.14\,^{+\,0.58}_{-\,0.13}$ & $0.14\,^{+\,0.58}_{-\,0.14}$ & $0.13\,^{+\,0.58}_{-\,0.13}$ & $0.13\,^{+\,0.57}_{-\,0.13}$ \\
\enddata
\end{deluxetable*}

\section{Discussion and Applications}
\label{sec:discusion}
%%%%%%%%%%%%%%%%%%%%%%%%%%%%%%%%%%%%%
%%%%%%%%%%%%%%%%%%%%%%%%%%%%%%%%%%%%%
%%%%%%%%%%%%%%%%%%%%%%%%%%%%%%%%%%%%%

There are several potential applications of our results to the interpretation of the abundances of now-$extinct$ SLRs that were present in the early Solar System (ESS) 4.6 Gyr ago, as well as the abundances of SLRs deposited in samples from the Earth \citep{wallner15,wallner16,feige18}, the Moon \citep{fimiani16}, and present in cosmic rays \citep{binns16}, all carrying $live$ SLR abundances in the Galaxy at the current time. Interpreting the abundances of the SLRs in the ESS provides us clues on the environment of the Solar System formation. This, for example, can shed light on the origin of \iso{26}Al ($\tau \simeq 1$ Myr) and help us determine if its high abundance in the ESS is typical for planetary systems in our Galaxy, which has implications on the thermo-mechanical evolution of planetesimals and the water content of terrestrial planets \citep{lichtenberg16b,lichtenberg19}. Also, by comparing the SLR abundances at the time of the ESS to their current abundances, we can derive information on the stellar sources of these nuclei, as done for example by \citet{hotokezaka15} and \cite{tsujimoto17} for the $r$ process, discussed in more detail below.

%%%%%%%%%%%%%%%%%%%%%%%%%%%%%%%%%%%
%%%%%%%%%%%%%%%%%%%%%%%%%%%%%%%%%%%
\subsection{General Considerations on the Determination of $\gamma$ and on the Isolation Time}
\label{sec:gamma}
%%%%%%%%%%%%%%%%%%%%%%%%%%%%%%%%%%%
%%%%%%%%%%%%%%%%%%%%%%%%%%%%%%%%%%%

The major problem in using our results effectively is that the parameter $\gamma$, the time interval between the formation of enrichment progenitors (see Table~\ref{tab:def}), is very uncertain and its values for different sources may range from 1\,Myr to a few hundred Myr. To determine from first principles the value of this parameter for each type of stellar source, we need to know (1) the overall rate of formation of each type of progenitor in the Galaxy, and (2) how many of these events were close enough to the parcel of gas in which the Solar System formed to contribute to its enrichment.

We can attempt to determine (1) from the initial mass function and the star formation rate. Alternatively, we can try to infer the formation of rate of different progenitors (i.e., $\gamma$) based on the observed rates of core-collapse supernovae, Type Ia supernovae, and AGB stars. Those observed rates, however, rather probe the enrichment events and therefore probe $\langle\delta\rangle$, the average time interval between two consecutive enrichment events. But as shown in Table~\ref{tab:delta}, $\langle\delta\rangle$ is similar to $\gamma$ in most cases, although the connection between these two parameters may be more complicated when accounting for an evolving star formation history (a temporal evolution for $\gamma$). For other types of events such as neutron star or neutron star-black hole mergers, on the other hand, the observational constraints are still very uncertain \citep{2017PhRvL.119p1101A}. More associated gravitational wave detections are required to better pinpoint their rate. It is even more difficult to establish (2) because we do not have yet a clear description of the transport and mixing of chemical species in the ISM from their creation event to the place where new stars are born. The resulting $\gamma$ can vary greatly depending on which scenario is considered \citep[see examples and discussion in Section~4.1 of][]{Lugaro2018}.

In spite of all these difficulties, it may be possible to use some SLRs to obtain clues on $\gamma$. This was attempted for the particularly interesting case of \iso{244}Pu to shed light on the debated astrophysical source of the $r$ process \citep{hotokezaka15,tsujimoto17}. It is possible to use the abundance of \iso{244}Pu to infer the Galactic rate of the $r$-process site for two reasons. First, this isotope has a relatively long mean-life of 115 Myr, which makes it relatively insensitive to a potential $isolation$ time (T$_\mathrm{iso}$) in relation to its ESS abundance. This isolation time refers to the total time that elapsed from the injection of the SLR into a given parcel of hot ISM gas to the time when the first solids (the calcium-aluminium inclusions, CAIs) formed in the ESS\footnote{The time of the formation of CAIs is the time at which the ESS SLR abundances are reported in the literature.}. Within a galactic context, the isolation time includes: (1) the time it took for the hot ISM matter to cool and form the molecular cloud where the Sun was born and (2) the time it took for the stellar cluster where the Sun was born to form. This definition does not include the further following three time intervals because they are much shorter than the (1)+(2) time interval: (i) the time it takes to transport the SLR from the creation event to the location of the formation of the new stars (considering, for example, ejecta velocities of 1000 km/s it would take 0.01 Myr to expand for 10 pc), (ii) the time it takes to form stars within a cluster, \citep[less than a Myr,][]{dib13} and (iii) the time interval for the formation of CAIs \citep[also of less than a Myr,][]{connelly12}\footnote{Note that, within a star-forming region, potential contributions from massive stars into the gas from which new stars are born could also contribute to the SLR abundances. In this case the relevant time intervals need to be modified to take into account, for example, the time that elapses between the birth of different stellar populations within the same star-forming region, and/or the time it may take for material from a particular nearby star or supernova to be injected into a collapsing cloud \citep[see, e.g.,][]{boss17,krause18}. In the case of the actinides such as \iso{244}Pu, however, this scenario is not relevant as we do not expect rare $r$-process sources to be present within a star-forming region.}. 
The isolation time for the Solar System is not known, and it could vary from a few Myr, if the Sun was formed in a small group \citep{hartmann01}, to up to 40 Myr, which is the 
observed lifetime of giant molecular clouds \citep{murray11}. This mean that at most 30\% of the abundance of \iso{244}Pu in the hot ISM would have decayed before the formation of the Solar System solids, and we can consider its ESS abundance representative within 30\% of its hot ISM abundance. 

The second reason why it is possible to use the abundance of \iso{244}Pu to infer the Galactic rate of the $r$-process site is that its abundance in the ISM is available from observations at two different times: today, as inferred from the Earth samples \citep{wallner15}, and at the time of the formation of the Sun, as inferred from meteorites \citep{hudson89}. Only by comparing the ESS abundance to the current ISM abundance it is possible to infer that large fluctuations in the ISM abundance of \iso{244}Pu are required to cover both points in time. If these fluctuations are due to heterogeneities, such comparison provides us with a determination of $\gamma$ of the order of hundreds of Myr, and this result indicates that rare events are the site of the $r$ process in the Galaxy \citep{hotokezaka15}. However, it should be kept in mind that the abundance of \iso{244}Pu in the ESS is still not well determined \citep[within roughly a factor of two, see discussion in Section 3.6 of][]{Lugaro2018}, and that the results of \citet{hotokezaka15} depend on the specific assumption that the material from the $r$-process sites is transported in the ISM via diffusion only.

Because the value of the isolation time is not well known and is comparable to the half-life of most SLRs, it is difficult to derive information on $\gamma$ using the ESS value of the SLRs only, as done by \citet{bartos19} on the basis of \iso{247}Cm with a half-life of 15.6 Myr. In fact, a hot ISM abundance higher than that observed in the ESS, as predicted by events more frequent than neutron star mergers \citep{bartos19}, cannot be excluded because such abundance could be decayed down to the ESS value by considering an isolation time.

Finally, we note that if $\tau/\gamma \lesssim 1$, by comparing the ESS abundances to the abundance calculated by GCE models, we can only derive the time elapsed between the last nucleosynthetic enrichment event and the formation of CAIs, which includes (and thus can be longer than) the isolation time described above. On the other hand, when $\tau/\gamma$ is greater than unity, comparing the ESS abundances to the abundance calculated by GCE models provides insights into the value of the actual isolation time, and the uncertainty on such value can be obtained from our statistical framework, as described in the next section (see also discussion in \citealt{lugaro14} and \citealt{cote19}).

%%%%%%%%%%%%%%%%%%%%%%%%%%%%%%%%
%%%%%%%%%%%%%%%%%%%%%%%%%%%%%%%%
\subsection{Application of our Results to the Abundances of 
SLR in the ESS}
\label{sec:scenario}

\begin{deluxetable*}{ccccccccc}
\tablewidth{0pc}
\tablecaption{Example of calculations of the isolation time T$_{\text{iso}}$ and its error bars derived for selected SLRs in the regime $\tau/\gamma \gtrsim 3$. The uncertainty multiplication factors in Column 6 represent the errors relative to the normalised values reported in Table~\ref{tab:uncertainties_box} for the corresponding adopted values of $\tau/\gamma$ and $\gamma$ (Columns 4 and 5), respectively. This error is applied to the abundance ratio calculated at the time of the formation of the Sun using the GCE code of \citet{cote19}. The Min, Best, and Max labels represent the uncertainties in Milky Way models (see their Figure~6). All $\tau$, $\gamma$, and T$_{\text{iso}}$ are given in Myr. The isotopes are listed in the same order as discussed in the text.\label{tab:final}}
\tablehead{
\colhead{SLR} & \colhead{Source} & \colhead{$\tau$} & \colhead{$\gamma$} & \colhead{$\tau/\gamma$} & Uncertainty & & \colhead{T$_{\text{iso}}$} & \\
\colhead{} & \colhead{(process)} & & \colhead{adopted} & \colhead{adopted} & factors & Min & Best & Max 
}
\startdata
$^{107}$Pd & AGB ($s$) & 9.4 & 3.16 & 3.16 & 
%$0.89\,^{+0.80}_{-0.60}$ & 
$0.61,1.39$ & 10$\,^{+3}_{-5}$ & 13$\,^{+3}_{-5}$ & 22$\,^{+3}_{-5}$ \\
$^{182}$Hf & AGB ($s$) & 12.8 & 3.16 & 3.16 & 
$0.61,1.39$ & 15$\,^{+4}_{-6}$ & 19$\,^{+4}_{-6}$ & 30$\,^{+4}_{-6}$ \\
$^{60}$Fe & CCSN & 3.78 & 1 & 3.16 & %$0.91\,^{+0.75}_{-0.58}$ & 
$0.63,1.37$ & 10$\,^{+1}_{-2}$ & 11$\,^{+1}_{-2}$  & 15$\,^{+1}_{-2}$ \\
%$^{53}$Mn & SNIa & 5.40 & 3 & 1 & $0.36, 1.82$ & 18$\,^{+3}_{-6}$ & 19$\,^{+3}_{-6}$ & 23$\,^{+3}_{-6}$ \\ 
% if DTD is SDIa, the isolation times are: 18, 19, 23
%$^{53}$Mn & SNIa & 5.40 & 3.16 & 1 & $<1.69$ & $<21$ & $<22$ & $<26$ \\ 
$^{92}$Nb & SNIa ($p$) & 50.1 & 10 & 3.16 & 0.60, 1.39 & -14$\,^{+16}_{-26}$ & -2$\,^{+16}_{-26}$ & 33$\,^{+16}_{-26}$ \\
%$^{92}$Nb & SNIa($p$) & 50.1 & 5 & 10 & 0.78, 1.23 & $<$... & $<$... & 33$\,^{+10}_{-12}$ \\
$^{146}$Sm & SNIa ($p$)   & 98  & 10 & 10 &  
$0.78,1.22$ &
$28\,^{+19}_{-25}$ & $51\,^{+19}_{-25}$ & $117\,^{+19}_{-25}$ \\
$^{146}$Sm & SNIa ($p$)    & 149  & 10 & 15 &  
$0.81,1.18$ & $104\,^{+25}_{-31}$ & $138\,^{+25}_{-31}$ & $235\,^{+25}_{-31}$ \\
\enddata
\end{deluxetable*}

% below is the table roughly before revision / referee report
%$^{107}$Pd & AGB ($s$) & 9.4 & 10 & 1 & 
%$0.89\,^{+0.80}_{-0.60}$ & 
%$0.33,1.90$ & 14$\,^{+6}_{-10}$ & 17$\,^{+6}_{-10}$ & 25$\,^{+6}_{-10}$ \\
%$^{182}$Hf & AGB ($s$) & 12.8 & 10 & 1 & 
%$0.89\,^{+0.80}_{-0.60}$ & 
%$0.33,1.90$ & 18$\,^{+8}_{-14}$ & 22$\,^{+8}_{-14}$ & 33$\,^{+8}_{-14}$ \\
%$^{60}$Fe & CCSN & 3.78 & 3 & 1 & %$0.91\,^{+0.75}_{-0.58}$ & 
%$0.36,1.82$ &
%10$\,^{+2}_{-4}$ & 11$\,^{+2}_{-4}$  & 15$\,^{+2}_{-4}$ \\
%$^{53}$Mn & SNIa & 5.40 & 3 & 1 & $0.36,1.82$ & 18$\,^{+3}_{-6}$ & 19$\,^{+3}_{-6}$ & 23$\,^{+3}_{-6}$ \\ 
% if DTD is SDIa, the isolation times are: 18, 19, 23
%$^{53}$Mn & SNIa+CCSN & 5.40 & 1 & 5 & 0.33, 1.90 & 17$\,^{+4}_{-6}$ & 19$\,^{+4}_{-6}$ & 23$\,^{+4}_{-6}$ \\
%$^{92}$Nb & SNIa ($p$) & 50.1 & 15 & 3 & 0.63, 1.43 & -14$\,^{+18}_{-23}$ & -2$\,^{+18}_{-23}$ & 33$\,^{+18}_{-23}$ \\
%$^{92}$Nb & SNIa($p$) & 50.1 & 5 & 10 & 0.78, 1.23 & $<$... & $<$... & 33$\,^{+10}_{-12}$ \\
%$^{146}$Sm & SNIa ($p$)   & 98  & 15 & 10 &  
%$0.79,1.23$ &
%$28\,^{+20}_{-23}$ & $51\,^{+20}_{-23}$ & $117\,^{+20}_{-23}$ \\
%$^{146}$Sm & SNIa ($p$)    & 149  & 15 & 10 &  
%$0.79,1.23$ & $104\,^{+31}_{-35}$ & $138\,^{+31}_{-35}$ & $235\,^{+31}_{-35}$ \\
%$^{146}$Sm & SNIa($p$)   & 98  & 5 & 10 &  
%% $0.78,1.23$ & these are for tau/gamma = 10, they below are right in-between these and the ones for tau/gamma=31.6
%$0.83,1.17$ &
%$27\,^{+15}_{-18}$ & $50\,^{+15}_{-18}$ & $117\,^{+15}_{-18}$ \\
%$^{146}$Sm & SNIa($p$)    & 149  & 5 & 30 &  
%$0.88,1.12$ & $103\,^{+17}_{-19}$ & $137\,^{+17}_{-19}$ & $234\,^{+17}_{-19}$ \\

Keeping in mind the difficulty of determining $\gamma$, we discuss an example scenario of the potential implications of our study to the abundances of SLR in the ESS. In the discussion below, we will refer to $\gamma$ as the recurrent time between events, since in our framework this is equivalent to $ \langle \delta \rangle$  (see Table~\ref{tab:delta}). We restrict this initial assessment only to isotopes whose abundances are relatively well known in the ESS, as already presented in Tables~2 and 3 of \citet{cote19}. 

\subsubsection{Galactic Chemical Evolution Models}
In the next subsections, we use the GCE code \texttt{OMEGA+} (\citealt{cote19}) to derive error bars for the isolation time T$_\mathrm{iso}$ of the ESS, using different SLRs and combining the uncertainties derived in \cite{cote19} and in our present work. In this context, the isolation time is the time elapsed between the last enrichment event that synthesized the considered SLRs and the formation of the first solids in the ESS.

\texttt{OMEGA+}\footnote{\url{https://github.com/becot85/JINAPyCEE}} is a two-zone homogeneous chemical evolution code that accounts for galactic inflows and outflows, star formation and stellar feedback, and the chemical enrichment generated by various sources such as core-collapse supernovae, Type~Ia supernovae, compact binary mergers, and AGB stars. The two zones consist of a galaxy surrounded by a large gas reservoir (i.e., the circumgalactic medium), in such a way that isotopes ejected from the galaxy by outflows can eventually fall back into the galaxy at later times.

In \cite{cote19}, we predicted the general evolution of radioactive-to-stable abundance ratios $M_\mathrm{radio}/M_\mathrm{stable}$ (see Figure~\ref{fig:convergence_2}) in the interstellar medium of our Galaxy, assuming the continuous enrichment scenario (i.e., no stochasticity). The input parameters of our models were calibrated to reproduce several observational constraints for the Milky Way at present time, including the star formation rate, the core-collapse and Type~Ia supernova rates, the star-to-gas mass ratio, and the gas inflow rate. We also ensured that our models reached solar metallicity by the time the Sun formed. Because of the error bars in the observations used to calibrate our models, we generated three different GCE models with final properties that bracket the observational constraints. 

One of the models fits the middle value of the observational constraints, labeled \textit{Best}. The two others aimed to minimize and maximize the abundance of SLRs in the galactic gas at the time of the ESS, while still remaining within the error bars of the observational constraints. Those models are labeled \textit{Min} and \textit{Max}. To minimize $M_\mathrm{radio}/M_\mathrm{stable}$, we minimized the star formation rate at the time of the ESS, the amount of stable isotopes locked inside stellar remnants, and the amount of stable isotopes trapped outside the galaxy in the circumgalactic medium. To maximized $M_\mathrm{radio}/M_\mathrm{stable}$, we did the opposite. Each of our models (Min, Best, and Max) is a different representation of the continuous enrichment of SLRs that could have occurred within our Milky Way galaxy, given the uncertainties in the observations mentioned above. 

To derive T$_\mathrm{iso}$ from our GCE models, we calculated the time needed to decay the predicted $M_\mathrm{radio}/M_\mathrm{stable}$ ratio down to the ESS values, assuming no enrichment event during that time window. In the next subsections, in order to account for the stochastic nature of enrichment events, we apply the relative uncertainties derived in the present work to the abundance of SLRs predicted by the Min, Best, and Max GCE models. This generated a probability distribution function for the predicted abundance of SLRs in the ESS and enabled us to derive error bars for T$_\mathrm{iso}$ for each of those three GCE models. We refer to \cite{Lugaro2018} for the list of stable reference isotopes of the SLRs considered below.

\subsubsection{\iso{107}Pd, \iso{129}I, \iso{182}Hf, and \iso{247}Cm}
The first case we consider is the well determined SLRs produced exclusively by the $r$ process: \iso{129}I and \iso{247}Cm; and those produced by the $r$ and the $s$ processes: \iso{107}Pd and \iso{182}Hf. If we assume that the $r$ process is produced by rare events, as indicated by the analysis of \iso{244}Pu described above and by the coincidence of the neutron star merger gravitational wave event GW170817 \citep{2017PhRvL.119p1101A}, then we can hypothesise that the $\tau/\gamma\lesssim 0.3$ regime is applicable in the case of the $r$ process and its SLRs (excluding the very long-lived uranium and thorium isotopes). This means that the $r$-process ESS abundances are most likely the results of one event only, which occurred roughly $100-200$\,Myr before the formation of the CAIs \citep{lugaro14,Lugaro2018,cote19,bartos19}. 

Under this hypothesis, the ESS abundances of \iso{107}Pd and \iso{182}Hf, which are produced by both the $r$ and the $s$ processes, can only carry the signature of the $s$ process, since their half lives of 6.5 and 8.9 Myr, respectively, assure that more than 90\% of their abundances produced by the last $r$-process event have decayed before the formation of the CAIs. The $\gamma$ for the progenitors of $s$-process events can be derived by considering stars of initial mass between roughly 2 and 4\,M$_\odot$ as the stellar source. However, since we do not know how far the slow ($10-30$ km/s) AGB winds can transport material in the Galaxy, it is difficult to tell how many of the total Galactic events will reach a given parcel of gas. By considering a simple ``snowplow'' scenario based on \citet{meyer00}, \citet{Lugaro2018} derived a potential $\gamma$ of the order of 50 Myr. In this case, also for the $s$-process isotopes we would be in the regime $\tau/\gamma\lesssim 0.3$ and probably only one $s$-process event contributed to the abundances of \iso{107}Pd and \iso{182}Hf in the ESS. 

On the other hand, it is possible that a more efficient ISM transport process would lower the value of $\gamma$ and therefore increase the $\tau/\gamma$ ratio. In this potential case of Regime I, we calculated the error bars on the isolation time T$_{\text{iso}}$, as reported in Table~\ref{tab:final}. Note that the T$_{\text{iso}}$ values reported here for \iso{107}Pd and \iso{182}Hf differ from those reported in Table~2 of \citet{cote19} because in that paper we considered both the $s$- and the $r$-process production for these two isotopes in the Galaxy, while here we re-run the models including only the $s$-process production. For any of the three different settings representing the uncertainties in the evolution of the Galaxy \citep{cote19}, the T$_{\text{iso}}$ values for these two isotopes overlap with each other (see also Figure~\ref{fig:scenario}). Within our framework, although the error bars are different for \iso{107}Pd and \iso{182}Hf because these isotopes have different mean-lives, they are not independent as they are assumed to come from the same stellar source. An in-depth study of the evolution of isotopic ratios involving two radioactive isotopes ejected by the same source will be presented by A. Yag\"ue et al.\,(in preparation).
Note that the error bars on the T$_{\text{iso}}$ are constant when varying the setting representing the uncertainties in the evolution of the Galaxy because T$_{\text{iso}}$ is a function of the logarithm of the abundance on which the uncertainty is applied. 
%The error bars, on the other hand, is a linearly dependent on $\tau$, which means that larger $\tau$ values on one hand allow us to approach more confidently the $\tau/\gamma > 1$ regime and decrease the uncertainty factors in Column 6 of Table~\ref{tab:final}, on the other, they result in larger error bars on the final T$_{\text{iso}}$ values. 

\subsubsection{\iso{53}Mn and \iso{60}Fe}
For the SLR isotopes of supernova origin considered in Table 3 of \citet{cote19}, \iso{53}Mn and \iso{60}Fe, it is also not trivial to specify the value of $\gamma$. \iso{60}Fe originates almost exclusively from core-collapse supernovae (CCSNe) and electron capture supernovae \citep[e.g.,][]{jones19}. If we assume that these events have a relatively low $\gamma$, potentially of 1\,Myr, then the regime $\tau/\gamma\gtrsim3$ applies and the error on the abundance and therefore on T$_{\text{iso}}$ is reported in Table~\ref{tab:final}. 
%The calculation of \citet{vasileiadis13} for the heterogeneous evolution of giant molecular clouds indicates values of $ \langle \delta \rangle$ of the order of $\sim$\,3 Myr (see their Figure 1). 
%It remains to be seen if this would be similar for the Galaxy in general.
Interestingly, the range of T$_{\text{iso}}$ values derived from \iso{60}Fe overlap at 9-11 Myr with the lowest T$_{\text{iso}}$ values from the two $s$-process isotopes, when considering the minimum GCE model, and almost overlap at 12-13 Myr, when considering the best GCE model (Figure~\ref{fig:scenario}). Type Ia supernovae (SNeIa) could significantly contribute to the abundance of \iso{53}Mn at the time of the formation of the Sun \citep{seitenzahl13}. Since roughly one in five supernovae is a SNIa \citep{cappellaro97}, the $\gamma$ value used for \iso{60}Fe should be multiplied by five. This means that for this isotope $\tau/\gamma \sim 1$ and Regime I cannot be applied.

\subsubsection{\iso{92}Nb and \iso{146}Sm}
Finally, we consider the case of the $p$-process isotopes \iso{92}Nb and \iso{146}Sm. These isotopes have the longest half lives of all SLRs well known in the ESS and if they are produced by supernovae, they will likely follow the $\tau/\gamma\gtrsim2$ regime. We test the scenario where they are produced exclusively by single degenerate SNeIa \citep{travaglio14}, while noting that \iso{92}Nb probably also originates from CCSNe \citep{travaglio18}. Assuming that single degenerate SNeIa represent 50\% of all SNeIa \citep{seitenzahl13}, the appropriate $\gamma$ to apply for consistency with \iso{60}Fe is of the order of 10 Myr. The resulting adopted $\tau/\gamma$ vary from 3.16 for \iso{92}Nb to 10-15 for \iso{146}Sm. The value for $\tau/\gamma$ of 15 was linearly interpolated from Table~\ref{tab:uncertainties_box} between $\tau/\gamma$ of 10 and 31.6, using $\log\tau$ as independent variable.

\begin{figure}
%\center
\includegraphics[width=3.37in]{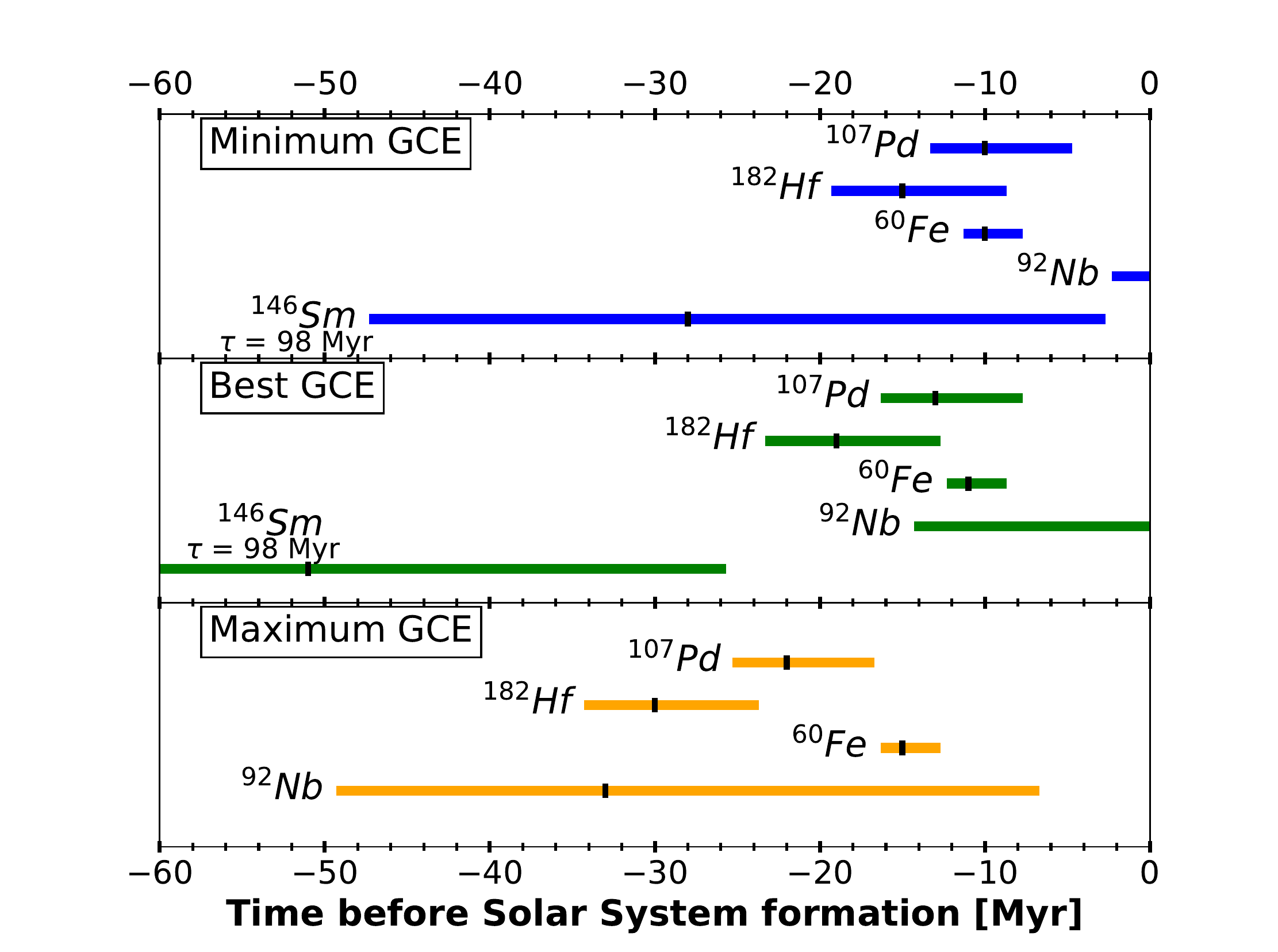}
\caption{Visual representation of the isolation times reported in Table~\ref{tab:final}, where the three different panels represent the three different Milky Way models presented in \cite{cote19}. The times derived from \iso{146}Sm when $\tau=149$\,Myr are out of scale.}
\label{fig:scenario}
\end{figure}

For \iso{92}Nb it is possible to derive a positive T$_{\text{iso}}$ value from homogeneous GCE only in the case when the GCE uncertainties are consider to provide the maximum value \citep[see Table 3 of][]{cote19}, in the other cases only upper limits can be given when considering the uncertainty factors derived here (Figure~\ref{fig:scenario}). For \iso{146}Sm we consider both current estimates of its half-life \citep{kinoshita12,marks14}. Because T$_{\text{iso}}$ depends linearly on $\tau$, the larger the $\tau$, the larger the errors on T$_{\text{iso}}$. The resulting values overlap with those of the other isotopes discussed above except for \iso{92}Nb only when we use the shortest value of the half-life of \iso{146}Sm and the minimum GCE model. Overall, the application of a possible scenario to the two $p$-process isotopes is problematic as it is not possible to derive times consistent between the two, as instead it is possible in the case of the two $s$-process isotopes \iso{107}Pd and \iso{182}Hf, as well in the case of the two $r$-process isotopes (C\^ot\'e et al.\,in preparation). More investigation is needed to understand the origin of these $p$-process nuclei and the half-life of \iso{146}Sm.

The hypothetical scenarios described above represent a simple test exercise to demonstrate the application of our results to SLRs, since several assumptions are invoked in the derivation of the T$_{\text{iso}}$ values and we have not considered the uncertainties in the stellar yields. It should also be noted that we assumed that no significant abundance of these isotopes is produced within the molecular cloud where the Sun was born by short-lived massive stars, which assumes that most of the ESS abundances can be derived from GCE. However, if such contribution existed to any of the SLRs discussed here, the isolation times corresponding to such isotopes could still be used as lower limits.

\subsubsection{\iso{26}Al, \iso{36}Cl, \iso{41}Ca, and \iso{235}U}
Finally, for \iso{26}Al, all regimes are possible, however, it is well known that the high ESS abundance of this nucleus cannot be reconciled with GCE (\citealt{huss09,cote19}) and many scenarios that involve the local molecular clouds and/or stellar cluster environment have been put forward to understand its origin \citep[e.g.,][]{gounelle12,pan12,young14,vasileiadis13,lichtenberg16a,boss17}. \iso{36}Cl and \iso{41}Ca, which were also present in the ESS, have such short mean-lives of 0.43 and 0.14\,Myr, respectively that they most likely fall in the $\tau/\gamma\lesssim 0.3$ regime and only probe the last enrichment event. The last cases are the very long-lived $r$-process isotopes of U and Th, which are important for the radiogenic heating of exoplanets \citep{uberseder14,frank14}. Because of their very long half lives, they cannot be used to measure T$_{\text{iso}}$. The effect of stocastic chemical evolution has less effect on these isotopes, but it could be important for \iso{235}U, which has a mean-life of 1\,Gyr. As compared to the potential $\gamma$ of the $r$-process events discussed above in the range $100-500$\,Myr, $\tau/\gamma$ could vary between 2 and 10 and Regimes I or II could apply. However, this case belongs to the class for which we cannot derive statistical properties because the system has not yet reached a steady state (empty cells in Table~\ref{tab:uncertainties_box}). For these long-lived isotopes, a specific, future study is required.

%%%%%%%%%%%%%%%%%%%%
%%%%%%%%%%%%%%%%%%%%
%%%%%%%%%%%%%%%%%%%%
\section{Conclusion}
\label{sec:conclusion}
%%%%%%%%%%%%%%%%%%%%
%%%%%%%%%%%%%%%%%%%%
%%%%%%%%%%%%%%%%%%%%

Using Monte Carlo calculations and analytical developments, we addressed the interplay between two characteristic timescales involved in the evolution of SLRs in our Galaxy: the SLR mean-life $\tau$ and the time interval $\gamma$ between the formation of the progenitor stars that will eventually produce nucleosynthetic events. We randomised the enrichment process in time (i.e., the intervals, $\delta$, between subsequent enrichment events) by randomly selecting the delay time between the birth of the progenitor star and the ejection of the SLR (see Figure~\ref{fig:param}), using different DTD functions relevant for different nucleosynthetic sources, from supernovae to neutron star mergers. All enrichment events are assumed to add the same amount of radioactive material in a given parcel of galactic gas. Our statistical approach allowed to quantify the uncertainty (68\,\% and 95\,\% confidence levels) in the predicted abundance of SLRs in the ISM, for different $\tau$, $\gamma$, and DTD functions.
%%%%%%applied a Monte Carlo method to analyse numerically the statistical properties of the resulting event distribution in time and of the abundance of the SLR, and tested the results analytically when possible.

Within our simplified but general framework, we found that the evolution of SLRs in a given parcel of gas can be categorized in terms of $\tau/\gamma$ (see Figures~\ref{fig:regime} and \ref{fig:pdf_all}). When $\tau/\gamma \gtrsim 2$ (Regime I), the spread around the median, which is in this case equal to $\tau/\gamma$, is symmetric with a standard deviation that can be approximated by Equation~(\ref{eq:absSpread}). The uncertainty in the predicted SLR abundances due to temporal heterogeneities in the built up of radioactive matter in the ISM, increases with decreasing $\tau/\gamma$ and reaches a maximum of 60\% at $\tau/\gamma=1$. When $\tau/\gamma \lesssim 1$, it is not possible to define a statistical distribution because the SLR abundance only carries the contribution of one or a few nucleosynthetic events. It is therefore more useful to calculate the probability that the observed SLR abundance carries the signature of one event only. If $\tau/\gamma < 0.3$ (Regime III), such probability is typically greater than 50\,\%, while the probability drops to zero when $\tau/\gamma$ increases to $\sim1$ (see Figure~\ref{fig:one_event}). When $0.3 \lesssim \tau/\gamma \lesssim 2$ (Regime II), the distribution is strongly asymmetric and the abundance is contributed by a small number of events.

For DTD functions that have a time frame larger than about 1\,Gyr, which is relevant for Type~Ia supernovae, AGB stars, and neutron star mergers, the spread of the abundance distribution of SLRs is independent of $\gamma$, regardless of the adopted $\tau/\gamma$ ratio. On the other hand, for short DTD functions that only span about 50\,Myr, which is relevant for winds and supernova explosions from massive stars, the spread depends on $\gamma$, except when $\tau/\gamma \lesssim 0.1$. Those results are visualized in Figure~\ref{fig:pdf_all}.

Although the value of $\gamma$ is still not well known for any given enrichment source (see Section~\ref{sec:gamma} for a discussion), we experimented by applying our results to analyse the implications of the stochastic enrichment process in the derivation of the isolation time of the Solar System from the ISM, using the abundance of different SLRs inferred to been present at the time of the ESS. We found that an isolation time between $9$ and 13\,Myr is consistent with the abundances of \iso{60}Fe, \iso{107}Pd, and \iso{182}Hf, given the uncertainties in the isolation times recovered using each of these three isotopes individually, under the assumption that Regime I is valid for these isotopes.

Future studies on the transport of matter from the site of production to the site of the birth of new stars in the ISM are required to constrain the value of $\gamma$ for different sources. Future work also needs to analyse the impact of varying the parameter $\gamma$. In this work, we assumed $\gamma$ to be constant, but its value should be varied with time because of its connection to the star formation history. It should also be varied with space, because star formation is not distributed homogeneously across the Galaxy. In an upcoming study, we plan to extend our framework into a 3D context, accounting for new parameters such as the distance of the source from the consider parcel of gas. Quantifying the impact of those additional effects will complement the impact of the time randomisation studied in the present work.

%%%%%%%%%%%%%%%%
%%%%%%%%%%%%%%%%
\acknowledgments
%%%%%%%%%%%%%%%%
%%%%%%%%%%%%%%%%
We thank M\'aria Pet\H{o} for discussions. We thank the anonymous referee for her/his comments, which significantly helped to improved the paper. This research is supported by the ERC Consolidator Grant (Hungary) funding scheme (Project RADIOSTAR, G.A. n. 724560). BC acknowledges the support from the National Science Foundation (NSF,
USA) under grant No. PHY-1430152 (JINA Center for the Evolution of the
Elements).

%\software{}

\vspace{5mm}
%\facilities{HST(STIS), Swift(XRT and UVOT), AAVSO, CTIO:1.3m,
%CTIO:1.5m,CXO}

\bibliographystyle{yahapj}
\bibliography{apj-jour,ms}

%% Include this line if you are using the \added, \replaced, \deleted
%% commands to see a summary list of all changes at the end of the article.
%\listofchanges

%% Appendix
\appendix

\section{Calculation of $\langle N_\mathrm{SLR}\rangle$ and $\langle N_\mathrm{SLR}^2\rangle$}
\label{sec:longMaths}

We can calculate $\langle N_\mathrm{SLR} \rangle$ by starting with the definition of the average value applied to Equation~(\ref{eq:evRadio}) from $t = 0$ to an arbitrary final time $t = t_f$
\begin{equation}
        \langle N_\mathrm{SLR} \rangle = \frac{1}{t_f}\int_0^{t_f}\left( e^{-t/\tau} + \sum_{j = 0}^{n - 1} e^{-(t - \sum_{i = 0}^j\delta_i)/\tau}\right) dt,
\end{equation}
we introduce the integral into the summation, but we must be careful to note that the $j_{th}$ term of $N_\mathrm{SLR}$ is only defined if $t \ge \sum_{i = 0}^j \delta_i$, which means that
\begin{equation}
        \langle N_\mathrm{SLR} \rangle = \frac{1}{t_f}\left(\int_0^{t_f} e^{-t/\tau} dt + \sum_{j = 0}^{n - 1} \int_{\sum_{i= 0}^j \delta_i}^{t_f}e^{-(t - \sum_{i = 0}^j\delta_i)/\tau} dt\right).
\end{equation}
Solving the integrals yields the expression
\begin{equation}
        \langle N_\mathrm{SLR} \rangle = \frac{\tau}{t_f} \left[\left(1 - e^{-t_f/\tau}\right) + \sum_{j = 0}^{n - 1} \left(1 - e^{-(t_f - \sum_{i = 0}^j \delta_i)/\tau}\right)\right].
\end{equation}
By taking $t_f = \sum_{i = 0}^n \delta_i + \Delta t = (n + 1) \langle \delta \rangle + \Delta t$, where $\Delta t < \delta_{n + 1}$, we can re-write this expression as
\begin{equation}
        \langle N_\mathrm{SLR} \rangle = \frac{\tau}{\langle \delta \rangle + \frac{\Delta t}{n + 1}} - \frac{\tau}{(n + 1) \langle \delta \rangle + \Delta t}\sum_{j = 0}^n e^{-(\Delta t + \sum_{i = j}^n \delta_i)/\tau}.
\end{equation}
In the approximation of $n \gg \tau/\langle \delta \rangle$ (equilibrium regime) and $n \langle \delta \rangle \gg \Delta t$ (large enough number of events), the average takes the value
\begin{equation}
    \langle N_\mathrm{SLR} \rangle \approx \frac{\tau}{\langle \delta \rangle}.
        \label{eq:avgProof}
\end{equation}

We can calculate $\langle N_\mathrm{SLR}^2 \rangle$ by the same procedure as before. We define first $N_\mathrm{SLR}$ in a more convenient way:
\begin{equation}
        N_\mathrm{SLR} = a_n + a_{n - 1} + a_{n - 2} + ... + a_0,
\end{equation}
where $a_{j > 0} = e^{-(t - \sum_{i = 0}^{j - 1} \delta_i)/\tau}$ and $a_0 = e^{-t/\tau}$. By squaring that sum and putting together the terms with the largest $n$, we have
\begin{equation}
        N_\mathrm{SLR}^2 = a_n^2 + 2a_n(a_{n - 1} + ... + a_0) + a_{n - 1}^2 + 2a_{n - 1}(a_{n - 2} + ... + a_0) + ... + a_0^2,
\end{equation}
and the average can be calculated with
\begin{equation}
        \langle N_\mathrm{SLR}^2 \rangle = \frac{1}{t_f} \int_0^{t_f} a_n^2 + 2a_n(a_{n - 1} + ... + a_0) + a_{n - 1}^2 + 2a_{n - 1}(a_{n - 2} + ... + a_0) + ... + a_0^2 \, dt.
\end{equation}
Just like in the calculation of the average value, we must integrate each term individually by taking into account the time from which they are defined. Therefore, the integral becomes
\begin{equation}
        \langle N_\mathrm{SLR}^2 \rangle = \frac{1}{t_f}\left[\int_{\sum_{i = 0}^{n - 1}\delta_i}^{t_f} a_n^2 + 2a_n(a_{n - 1} + ... + a_0) \, dt + \int_{\sum_{i = 0}^{n - 2}\delta_i}^{t_f}a_{n - 1}^2 + 2a_{n - 1}(a_{n - 2} + ... + a_0) \, dt + ... + \int_0^{t_f} a_0^2 \, dt \right].
\end{equation}
The integral leading with the term $a_j^2$ can be expressed as
\begin{equation}
        I_j = \int_{\sum_{i = 0}^{j - 1}\delta_i}^{t_f} e^{-2(t - \sum_{i = 0}^{j - 1}\delta_i)/\tau} + 2e^{-(2t - \sum_{i = 0}^{j - 1}\delta_i - \sum_{i = 0}^{j - 2}\delta_i)/\tau} + 2e^{-(2t - \sum_{i = 0}^{j - 1}\delta_i - \sum_{i = 0}^{j - 3}\delta_i)/\tau} + ... + 2e^{-(2t - \sum_{i = 0}^{j - 1}\delta_i)/\tau}\, dt,
\end{equation}
for which the solution, with $t_f = \sum_0^n \delta_i + \Delta t$ as before, is
\begin{equation}
        I_j = \frac{\tau}{2}\left\{\left[1 - e^{-2(\sum_{i = j}^{n}\delta_i + \Delta t)/\tau}\right] + ... + 2 \left[e^{-\sum_{i = k}^{j - 1}\delta_i/\tau} - e^{-(2\sum_{i = j}^n\delta_i + \sum_{i = k}^{j - 1}\delta_i + 2\Delta t)/\tau}\right] + ... \right\},
\end{equation}
where the index $k$ starts at $k = j - 1$ and goes down to $k = 0$. In a more compact way, and by extracting the common factors, the term $I_j$ becomes
\begin{equation}
        I_j = \frac{\tau}{2}\left[1 - e^{-2(\sum_{i = j}^{n}\delta_i + \Delta t)/\tau}\right]\left[1 + 2\sum_{k = 0}^{j - 1} e^{-\sum_{i = k}^{j - 1}\delta_i/\tau}\right].
\end{equation}
And the expression for the average of the square of $N_\mathrm{SLR}$ becomes
\begin{equation}
        \langle N_\mathrm{SLR}^2 \rangle = \frac{\tau}{2t_f}\sum_{j = 0}^n \left\{\left[1 - e^{-2(\sum_{i = j}^{n}\delta_i + \Delta t)/\tau}\right]\left[1 + 2\sum_{k = 0}^{j - 1} e^{-\sum_{i = k}^{j - 1}\delta_i/\tau}\right]\right\}.
\end{equation}

In order to analyze the previous expression, we consider from now on that we are on the equilibrium regime, with $n \gg \tau/\delta$, and that there are enough events such that
\begin{equation}
        \left[1 - e^{-2(\sum_{i = j}^{n}\delta_i + \Delta t)/\tau}\right] \approx 1
\end{equation}
for the majority of the $j$, leaving us with
\begin{equation}
        \langle N_\mathrm{SLR}^2 \rangle \approx \frac{\tau}{2t_f}\sum_{j = 0}^n \left[1 + 2\sum_{k = 0}^{j - 1} e^{-\sum_{i = k}^{j - 1}\delta_i/\tau}\right] = \frac{\tau}{2t_f}\sum_{j = 0}^n \left[1 + 2S_j\right] = \frac{\tau}{2\langle \delta \rangle}\left(1 + 2\langle S \rangle\right)
\end{equation}
Using this expression, we can calculate the exact value for the standard deviation using Eq. (\ref{eq:avgProof}) as
\begin{equation}
    \sigma = \sqrt{\frac{\tau}{2\langle \delta \rangle}}\sqrt{1 + 2\langle S \rangle - \frac{2\tau}{\langle \delta \rangle}}.
    \label{eq:sqAvgExact}
\end{equation}
From the definition of $\langle S \rangle$ we know that, in equilibrium,
\begin{equation}
    \langle S \rangle = \langle e^{-\delta/\tau} \rangle + \langle S e^{-\delta/\tau} \rangle,
\end{equation}
which allows us to obtain a more intuitive expression by taking the approximation
\begin{equation}
    \langle Se^{-\delta/\tau} \rangle \approx \langle S \rangle \langle e^{-\delta/\tau} \rangle,
\end{equation}
from where we can solve $\langle S \rangle$ as
\begin{equation}
    \langle S \rangle \approx \frac{\langle e^{-\delta/\tau} \rangle}{1 - \langle e^{-\delta/\tau}\rangle}.
\end{equation}
Using this expression in Eq. (\ref{eq:sqAvgExact}), gives us
\begin{equation}
    \sigma \approx \sqrt{\frac{\tau}{2\langle \delta \rangle}}\sqrt{\frac{1 + \langle e^{-\delta/\tau} \rangle}{1 - \langle e^{-\delta/\tau} \rangle} - \frac{2\tau}{\langle \delta \rangle}}.
    \label{eq:sqAvgProof2}
\end{equation}
Finally, for $\delta/\tau \ll 1$ we have that
\begin{equation}
    \langle e^{-\delta/\tau} \rangle \approx 1 - \frac{\langle \delta \rangle}{\tau} + \frac{\langle \delta^2 \rangle}{2 \tau^2},
\end{equation}
and therefore
\begin{equation}
    \frac{1 + \langle e^{-\delta/\tau} \rangle}{1 - \langle e^{-\delta/\tau} \rangle} \approx \frac{4\tau^2 - 2\tau\langle \delta \rangle + \langle \delta^2 \rangle}{2\tau\langle \delta \rangle - \langle \delta^2 \rangle},
\end{equation}
which when introduced in Equation~(\ref{eq:sqAvgProof2}) gives
\begin{equation}
    \sqrt{\frac{\tau}{2 \langle \sigma \rangle}}\sqrt{\frac{1 + \langle e^{-\delta/\tau} \rangle}{1 - \langle e^{-\delta/\tau} \rangle} - \frac{2\tau}{\langle \delta \rangle}} = \sqrt{\frac{\tau}{2 \langle \sigma \rangle}}\sqrt{\frac{2\tau(\langle \delta^2 \rangle - \langle \delta \rangle^2) + \langle \delta^2 \rangle \langle \delta \rangle}{2\tau\langle \delta \rangle^2 - \langle \delta^2 \rangle \langle \delta \rangle}}.
\end{equation}
When defining $\sigma_\delta^2 = \langle \delta^2 \rangle - \langle \delta \rangle^2$ and supposing that $\langle \delta^2 \rangle \langle \delta \rangle$ is smaller than the terms it's being added or subtracted from, we have that the standard deviation for $N_\mathrm{SLR}$ becomes
\begin{equation}
    \sigma \approx \frac{\sigma_\delta}{\langle \delta \rangle} \sqrt{\frac{\tau}{2\langle \delta \rangle}}
    \label{eq:sqAvgProof3}
\end{equation}

\end{document}